\documentclass[a4paper,11pt]{article}
\pdfoutput=1 

\usepackage{jcappub} 

\usepackage[T1]{fontenc} 
\usepackage{graphicx}
\usepackage{subcaption}

\def\fnl{f_{\mathrm{NL}}}

\title{\boldmath It's All About the Environment: Local $f_{\rm NL}$ from a Dark Matter Conditioned Multitracer Analysis}

\author[a, b]{Julius Adolff}
\author[a, b]{and Uroš Seljak}
\affiliation[a]{Department of Physics, University of California, Berkeley, CA 94720, USA}
\affiliation[b]{Lawrence Berkeley National Laboratory, University of California, Berkeley, CA 94720, USA}

\emailAdd{juliusadolff@berkeley.edu}

\abstract{Primordial non-Gaussianity (PNG) offers a powerful test of inflationary physics beyond the simplest single-field scenarios. In large-scale structure, biased tracers respond to local-type PNG with a distinctive scale-dependent signature that is strongest on large scales, but single-tracer measurements are fundamentally limited by cosmic variance. Multitracer methods can mitigate this limitation; however, many existing proposals require splitting tracers using a wide range of halo masses which are often not available, or using secondary halo properties that are not directly observable.
Here we develop and validate an alternative strategy based on splitting a tracer sample by its large-scale {dark matter} environment. We derive a theoretical framework for the $f_{\rm NL}$ response of environmentally selected tracers and test it in $N$-body simulations, confirming consistency between separate-universe predictions and direct power-spectrum inference. 
To enable observational implementation, we show that a simple linear-theory reconstruction of the dark matter environment from halos yields unbiased $f_{\rm NL}$ constraints in \texttt{Quijote} simulations.
Forecasts for the DESI LRG sample indicate that environmental multitracing can improve constraints on local PNG by a factor of $2$--$3$ relative to conventional single-tracer analyses for high density tracers, comparable to gains achieved when halo formation time is assumed to be perfectly known. We develop an analytic model explaining why the 
gains are smaller for a halo environment split and why separate-universe predictions fail in this case. 
Unlike approaches that require modeling assembly bias or inferring secondary halo properties, our method relies only on linear theory. These results establish environmental multitracer techniques as a promising route toward $\sigma(f_{\rm NL})\sim 1$ with upcoming galaxy surveys.
}

\begin{document}
\maketitle
\flushbottom

\section{Introduction}
\label{sec:intro}
One of the central goals of modern cosmology, and particularly the study of large-scale structure (LSS), is to deepen our understanding of the early Universe. While analyses of the Cosmic Microwave Background (CMB) have provided compelling evidence for the simplest models of slow-roll inflation, predicting a nearly scale-invariant primordial power spectrum and adiabatic, Gaussian initial conditions~\cite{Akrami_2020, Aghanim_2020}, theoretical and observational work continues to probe the details of this inflationary phase~\cite[see e.g.][for a review]{Baumann_2009}. Extending our understanding beyond the constraints provided by the CMB requires exploring how early-universe processes influence the late-time clustering of galaxies and other LSS tracers~\cite{Desjacques_2018, Liguori_2010, Dalal_2008}. As the data collected in galaxy surveys becomes increasingly vast and precise, sophisticated methodologies for extracting this information are becoming essential~\cite{DESI_2016}.

Of particular interest are observational tests that can distinguish between simple slow-roll models and more complex theories involving multiple fields or modifications to General Relativity at high energies~\cite{Maldacena_2003, Bartolo_2004}. The most common test probes deviations from Gaussianity in the initial conditions by expanding the primordial potential $\Phi$ in terms of a Gaussian field $\phi$ to second order:
\begin{equation}
    \label{eq:expansionFNL}
    \Phi(\boldsymbol x) = \phi(\boldsymbol x) + \fnl (\phi(\boldsymbol x)^2 - \langle\phi^2  \rangle),
\end{equation}
where $\fnl$ parameterizes (local) deviations from Gaussianity~\cite{Komatsu_2001}. Using Planck observations of the CMB bispectrum, $\fnl$ can be constrained to $-0.9 \pm 5.1$ \cite{planckcollaboration2019planck2018resultsix}. However, many compelling models, e.g. the curvaton scenario, require measurements of primordial non-Gaussianity at the $\sigma(f_{\rm NL}) \sim 1$ level~\cite{Alvarez_2014,dePutter_2014} in order to distinguish between different inflationary scenarios in a statistically informative way; achieving such precision from the CMB alone is limited by cosmic variance~\cite{dePutter_2014}. Instead, it has long been appreciated that primordial non-Gaussianity induces a unique scale-dependent signature in the clustering of galaxies, strongest on large scales~\cite{Dalal_2008, Matarrese_2008, Slosar_2008}. Extracting this signal is a primary goal of LSS surveys~\cite{Dore_2014, LSSTScienceBook_2009, Bock_2026}.

Observational efforts have so far focused on extracting this signal from the auto-power spectrum of a single tracer, with the tightest constraints obtained from the DESI quasar sample \cite{chaussidon2025constrainingprimordialnongaussianitydesi}.\footnote{Although an incremental improvement can be made by also including the cross spectrum with the DESI LRG sample \cite{chaussidon2025constrainingprimordialnongaussianitydesi}.} Such single-tracer analyses are, however, fundamentally limited by cosmic variance and are therefore known to be suboptimal. Cosmic variance cancellation can be achieved by considering multiple tracers with different biases \cite{Seljak_2009}. Several studies have explored the improvements obtainable by exploiting the dependence of the $\fnl$ response on secondary halo properties, such as concentration \cite{Lazeyras_2023, kvasiuk2024talefieldsneuralnetworkenhanced, Sullivan:2023qjr}, formation time \cite{Reid_2010, Slosar_2008, Fondi_2024} and size \cite{nguyen2026galaxysizescomplementaryzerobias}, and usually find that $\sigma(\fnl)$ can be improved by a factor of $2$--$3$ through cosmic variance cancellation.

Applying such methodologies to real data is challenging for a multitude of reasons. First, since the secondary halo properties on which the $\fnl$ response depends are not directly observable, one can only attempt to infer them indirectly from observational characteristics such as color and luminosity \cite{Sullivan:2023qjr, Fondi_2024} or higher order bias parameters \cite{SimoPNG}. Second, there has been some discussion that since galaxies do not trace halos, and the halo occupation distribution (HOD) itself depends on $\fnl$ \cite{Voivodic_2021}, care has to be taken when relating the $\fnl$ response of a halo to that of the galaxies it hosts. Furthermore, certain galaxy populations might preferentially reside in halos with secondary properties that do not match the average across the entire population, further complicating estimates of the $\fnl$ response \cite{fondi2026assemblybiaslocalprimordial, Barreira_2020}. Both challenges can in principle be addressed by measuring the time evolution of the number density of some tracer (defined, e.g., via a color cut) and thereby inferring the $\fnl$ response directly \cite{Sullivan:2025fie, dalal2025estimatingnongaussianbiasusing}. This remains observationally challenging since it requires accurately modeling the time evolution of the selection function \cite{Sullivan:2025fie}.

Here, we therefore propose an alternative method, where one splits the given tracer sample by its local dark matter environment to obtain two samples with very different linear bias but equal $\fnl$ response. Similar approaches have previously been proposed in the literature, in the context of zero-bias tracers \cite{Castorina_2018} and the density-split formalism \cite{morawetz2024constrainingprimordialnongaussianitydensitysplit}. In this work we develop a coherent theoretical framework, validated against simulations, that helps explain the mechanisms driving the environmental dependence of the $\fnl$ response and reconciles divergent results noted in previous work \cite{morawetz2024constrainingprimordialnongaussianitydensitysplit, Merino_2026}. Based on our formalism, we forecast that the constraints obtained from the DESI LRG sample can be improved by a factor of $2$--$3$ for high density tracers with our method. Crucially, our method relies only on linear theory. It requires neither the modeling of assembly bias nor the inference of secondary halo properties. It nevertheless attains constraining power comparable to methods that assume perfect knowledge of secondary properties such as the halo formation time.

This paper is structured as follows. In Section~\ref{sec:TheoBack} we review the theoretical formalism that predicts the $\fnl$ response of biased tracers and extend it to the case of environmental selection. In Section~\ref{sec:ValSim} we validate our description extensively against simulations and discuss why the halo environment is less informative than the dark matter environment. In Section~\ref{sec:Obs} we connect to observations and demonstrate that a simple linear-theory estimator for the dark matter environment offers similar improvements, while remaining unbiased. We conclude in Section~\ref{sec:Conc} with a discussion of the implications and future prospects.

\section{Theoretical Background}\label{sec:TheoBack}

\subsection{The Impact of $\fnl$ on Galaxy Clustering}
We briefly review how to compute the impact of primordial non-Gaussianity (PNG) on the clustering statistics of a tracer. In particular, we want to emphasize a key assumption that is made in the usual derivation. PNG induces a coupling between large range and short range scales, which breaks the symmetry arguments that predict a scale-independent linear bias~\cite{Dalal_2008, McDonald_2008}. Following the standard approach, we decompose the matter field into its long- and short-wavelength contributions:
\begin{equation}
    \delta_m = \delta_\ell + \delta_s .
\end{equation}
The presence of PNG does not modify $\delta_\ell$ on linear scales, since it only receives a contribution at quadratic order. Similarly decomposing the gravitational potential as $\phi = \phi_\ell + \phi_s$, the change in the short-range contribution from Eq.~\eqref{eq:expansionFNL} reads
\begin{equation}
    \delta_s \rightarrow \delta_s \left(1 + 2 \fnl \phi_\ell + \fnl \phi_s \right),
\end{equation}
since $\delta$ and $\phi$ are related linearly via the Poisson equation. The presence of non-zero PNG thus leads to a modulation of short-range clustering $\delta_s$, while leaving the large scales modes $\delta_\ell$ unchanged. This can be interpreted as a modulation of the local value of $\sigma_8$ with
\begin{equation}
    \sigma_8 \rightarrow (1 +  2\fnl \, \phi_\ell)\sigma_8 \Rightarrow \frac{\partial \log \sigma_8}{\partial (\fnl \phi_\ell)} = 2.
    \label{eq:Dsigma8}
\end{equation}
The linear bias of a tracer takes the form
\begin{align}
    b &= \frac{1}{\bar n}\frac{d\bar n}{d\delta_\ell} \nonumber\\
      &= \frac{1}{\bar n}\left( \frac{\partial \bar n}{\partial \delta_\ell}
      + \frac{\partial \phi_\ell}{\partial \delta_\ell}\frac{\partial \bar n}{\partial \phi_\ell} \right),
      \label{eq:bFULL}
\end{align}
where $\bar n$ is the tracer number density. To compute the dependence of the number density $\bar n$ on $\phi_\ell$ we can make use of Eq. \eqref{eq:Dsigma8} to relate the dependence on  $\phi_\ell$ to the dependence of $\bar n$ on $\sigma_8$. This leads to the usual result for the $\fnl$ response:
\begin{equation}
     b_\phi \equiv \frac{d \log \bar n}{d (\fnl \phi_\ell)} = 2 \frac{\partial \log \bar n}{\partial \log \sigma_8},
     \label{eq:SUb_phi}
\end{equation}
so that Eq.~(\ref{eq:bFULL}) becomes
\begin{equation}
    b = b_1 + \frac{\partial \phi_\ell}{\partial \delta_\ell}\, b_\phi\, \fnl .
    \label{eq:ScaleDepBias}
\end{equation}
We will later demonstrate that for halos selected by their own halo environment, Eq.~(\ref{eq:SUb_phi}) receives an additional contribution from the \textit{explicit} dependence of the number density $\bar n$ on the value of the gravitational potential $\phi_\ell$.

To compute $\partial \phi_\ell/\partial \delta_\ell$, we use the relation between the gravitational potential and the matter overdensity in Fourier space \cite{Dalal_2008}
\begin{equation}
    \delta_m(k) = \mathcal M(k)\phi(k), \qquad 
    \mathcal M(k) = \frac{2}{3}\frac{D(z)\,k^2 T(k)}{\Omega_m H_0^2}.
\end{equation}
Equation~(\ref{eq:ScaleDepBias}) then implies (on linear scales)
\begin{equation}
    \delta_t(k) = \left(b_1 + b_\phi\,\frac{\fnl}{\mathcal M(k)} \right)\delta_m(k) + \epsilon,
\end{equation}
where $\epsilon$ is the stochastic contribution to the tracer field. Since $\mathcal M(k)\sim k^2$, the PNG correction to the bias is strongest on large scales and can therefore be understood within linear theory. The value of $b_\phi$, however, depends on the full non-linear formation process of the tracer and its dependence on astrophysical parameters is still an active area of research \cite{Barreira_2020, Barreira_2022, perez2026impactgalaxyformationgalaxy}.

In this work we focus exclusively on mass-selected halos, for which the universality relation \cite{Dalal_2008, Slosar_2008}
\begin{equation}
    b_\phi = 2\delta_c\,(b_1-1)
\end{equation}
is a robust prediction validated in $N$-body simulations \cite{Desjacques_2009, Desjacques_2010}. As we will show, our method is insensitive to the specific value of $b_\phi$ and can be reformulated entirely in terms of constraints on $b_\phi\,\fnl$. Simulation or theory-based priors on $b_\phi$ can then be used to extract the posterior of $\fnl$.

\subsection{The Dependence on the Environment}
As noted previously, the $\fnl$ response of halos depends on many secondary halo properties, such as concentration and formation time. When considering a given population of halos, measuring $\fnl$ from the power spectrum of the full population is therefore suboptimal, since it averages over the intrinsic variation of $b_\phi$ across halos. More optimal analyses seek to split (or weigh) the population using a proxy that is sensitive to $b_\phi$.

Motivated by previous studies \cite{Castorina_2018, morawetz2024constrainingprimordialnongaussianitydensitysplit}, we propose splitting the data according to a quantity that is easily inferred from observations: its large-scale dark matter environment. As we will derive shortly, if we split a tracer into two equal-sized bins based on its dark-matter environment, then the $\fnl$ response of each subpopulation is exactly equal to that of the total sample. This has several desirable implications. First, it means that the method works independently of how $b_\phi$ is computed and can be used to constrain $b_\phi\,\fnl$. Second, and most importantly, since $b_1$ can change drastically across the two subsamples, the ratio $b_1/b_\phi$ also changes. For multitracer analyses, the Fisher information on $\fnl$ scales as \cite{Seljak_2009, barreira2023optimalrobustfrmnl}
\begin{align}
    F_{\fnl} &\propto V\int dk\,k^2W(k)\left( \frac{b_1^{(1)}b_\phi^{(2)} - b_1^{(2)}b_\phi^{(1)}}{\left(b_1^{(1)}+\fnl b_\phi^{(1)}/\mathcal M(k)\right)\left(b_1^{(2)}+\fnl b_\phi^{(2)}/\mathcal M(k) \right)} \right)^2 \\
    &\propto \left(b_\phi^{(1)}b_\phi^{(2)}\right)^2\left( b_1^{(1)}/b_\phi^{(1)} - b_1^{(2)}/b_\phi^{(2)} \right)^2 ,\label{eq:ScalingLaw}
\end{align}
where we have taken the cosmic variance limit.\footnote{ The per-mode weight function is found to be
\begin{equation}
    W(k) = \frac{1}{X^{(1)}(k) + X^{(2)}(k)} \frac{P_{mm}(k)}{\mathcal M(k)^{2}},
\end{equation}
with $X^{(i)}(k) = \left(b_1 ^{(i)} + \fnl b_\phi^{(i)}/\mathcal M(k) \right)^{-2}/\bar n^{(i)}$, following the notation of \cite{barreira2023optimalrobustfrmnl}.
}
From Eq.~\eqref{eq:ScalingLaw} we can conclude that large changes in $b_1/b_\phi$ between the two tracers can yield substantial gains even when $b_\phi$
itself does not change. In this case, the above reduces to ${F_{\fnl} \propto \left(b_1^{(1)} - b_1^{(2)}\right)^2}$.

To explain why the $\fnl$ response of a halo population remains unchanged when the tracer is split into two equal-sized subpopulations, we need to compute how $b_\phi$ depends on the large-scale dark matter environment. Let $(\delta_m)_R$ denote the matter field smoothed on a radius $R$. We use, throughout this work,
$R = 10\,\mathrm{Mpc}/h$ and a Gaussian smoothing filter.\footnote{Many of our findings are robust to changes in the smoothing radius. However, if $R$ is chosen too small, the value of $(\delta_m)_R$ at halo positions becomes increasingly noisy and less informative about the large-scale environment. If $R$ is chosen too large, then different environments are averaged over and $b_\phi$ becomes less sensitive to $(\delta_m)_R$. Moreover, the smoothing radius sets the range of accessible scales, $k \lesssim 1/R$. We found $R = 10\,\mathrm{Mpc}/h$ to be a reasonable balance between these extremes; nonetheless, we verified that many quoted numerical values change only slightly when $R$ is varied.}
Selecting halos by their environment modulates the number density of the parent population in a simple manner. Of the full sample, we keep only the fraction whose smoothed environment lies in an infinitesimal range around $\delta$. Writing $p(\delta)$ for the probability that a halo's environment equals $\delta$, the number density of this selected subpopulation is therefore the parent number density $\bar n$ multiplied by that probability,
\begin{equation}
    \bar n_{\rm sel} = p(\delta)\,\bar n .
\end{equation}
Since $b_\phi$ measures the response of the abundance to a long-wavelength change in $\sigma_8$, the environment dependence of $b_\phi$ is inherited entirely from that of $p(\delta)$.

Using Eq.~(\ref{eq:SUb_phi}), the $\fnl$ response of this subpopulation is
\begin{equation}
    b_\phi(\delta) =
    \underbrace{2\frac{\partial \log \bar n}{\partial \log \sigma_8}}_{=b_\phi}
    +
    \underbrace{2\frac{\partial \log p(\delta)}{\partial \log \sigma_8}}_{=\Delta b_\phi(\delta)}.
\end{equation}
Computing $\Delta b_\phi(\delta)$ requires specifying the probability $p(\delta)$. This cannot be obtained directly from the probability distribution of the total dark matter field, since we only evaluate the field at halo positions. Instead, we require the distribution of dark matter conditioned on the fact that a halo formed at the given position. In Appendix~\ref{app:CondDist} we argue that this conditional distribution is well approximated as a Gaussian
\begin{equation}
    (\delta_m)_R \,|\,\text{halo formed} \sim \mathcal N(\mu,\sigma^2),
\end{equation}
where $\mu$ is independent of $\sigma_8$ and $\sigma \propto \sigma_8$.
Intuitively, the environment of a halo is set by the large-scale matter fluctuations around it, which are close to Gaussian. The mean $\mu$ is related to the collapse threshold $\delta_c$ and therefore independent of $\sigma_8$, whereas the width $\sigma$ is determined by the size of the environment fluctuations and hence scales linearly with $\sigma_8$. We derive both properties in Appendix~\ref{app:CondDist}. The probability $p(\delta)$ is then
\begin{equation}
    p(\delta) = \frac{1}{\sigma\sqrt{2\pi}}
    \exp\left[-\frac{(\delta-\mu)^2}{2\sigma^2}\right]\,d\delta,
\end{equation}
which implies
\begin{equation}
    \frac{\partial \log p(\delta)}{\partial \log \sigma}
    = \left(\frac{\delta-\mu}{\sigma}\right)^2 - 1.
\end{equation}
Since $\sigma \propto \sigma_8$, we have $d\log\sigma = d\log\sigma_8$, and therefore the dependence of $b_\phi$ on the dark matter environment is
\begin{equation}
    b_\phi\!\left[(\delta_m)_R\right]
    = b_\phi
    + 2\left[\left(\frac{(\delta_m)_R-\mu}{\sigma}\right)^2 - 1\right].
    \label{eq:b_phi_delta_m}
\end{equation}
Crucially, this relation and the probability $p(\delta)$ are even around $(\delta_m)_R-\mu$. Therefore, if a sample is split into two equal-sized subpopulations,
$(\delta_m)_R \le \mu$ and $(\delta_m)_R > \mu$, the value of 
\begin{equation}
    b_\phi^{\rm sel} = b_\phi + \frac{\int_{\delta\, \in\, \text{(given bin)}}    p(\delta )\, \Delta b_\phi(\delta )}{\int_{\delta\, \in\, \text{(given bin)}}    p(\delta )}
\end{equation}
must be the same for both. Since the two bins have equal weight and their
sum reproduces the full-sample response $b_\phi$, this implies that each
subpopulation has $\fnl$ response equal to $b_\phi$, as stated above. 

\begin{figure}[t]
    \centering
    \includegraphics[width=0.49\linewidth]{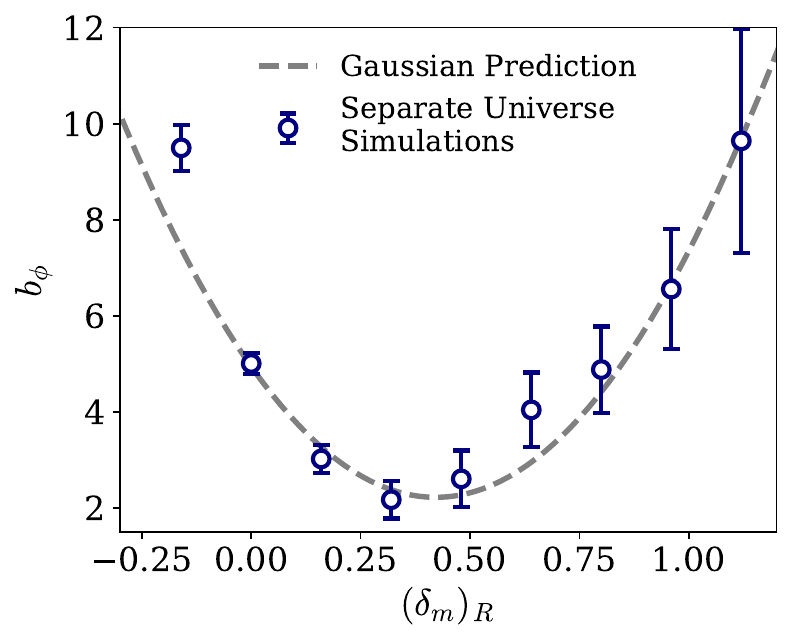}
    \hfill
    \includegraphics[width=0.49\linewidth]{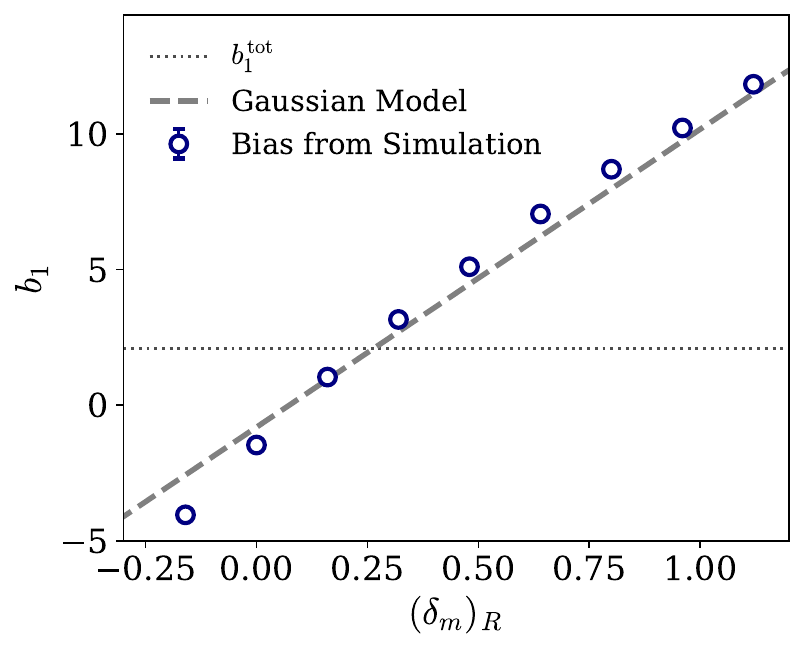}
    \caption{Comparison of the theoretical dependence of the bias parameters $b_1$ and $b_\phi$ on the large-scale dark matter environment under the Gaussian model (gray, dashed) to the results from $N$-body simulations (blue dots). The left plot is based on 20 \texttt{FastPM} (10 seed) runs \cite{Feng_2016}, used to compute the derivative in Eq.~(\ref{eq:SUb_phi}) via finite differences. The right plot shows the dependence of $b_1$ on the smoothed dark matter environment, computed from the \texttt{Halfdome} simulations \cite{Bayer_2025} using the estimator in Eq.~(\ref{eq:Estimator1}) on scales $k < 0.015\,h/\mathrm{Mpc}$. Both plots consider halos in the mass bin ${M \in [1.5\cdot 10^{13}, 4.64\cdot 10^{14}]\,M_\odot/h}$ at redshift $z=0.5$. Details of the simulations are discussed in Section~\ref{sec:ValSim}. The theoretical predictions are discussed in Section~\ref{sec:TheoBack}.}
    \label{fig:DarkMatterSplit}
\end{figure}

Intuitively, one expects tracers in high dark matter density environments to have a higher linear bias than the average, and conversely for regions with low environment. In the present framework, this can be derived as follows. A long-wavelength perturbation $\delta_\ell$ shifts the mean $\mu \rightarrow \mu + \delta_\ell$. The change in the linear bias is therefore
\begin{equation}
    \Delta b_1
    = \frac{\partial \log p}{\partial \mu}
    = \frac{(\delta_m)_R-\mu}{\sigma^2}
    \quad\Rightarrow\quad
    b_1\left[(\delta_m)_R\right] \propto (\delta_m)_R.
    \label{eq:KaiserBias}
\end{equation}
This result, that the linear bias is proportional to the local dark matter environment, is well known in the literature \cite{Pinon_2025, 1984ApJ} and sometimes referred to as Kaiser bias. We note that, unlike Eq.~(\ref{eq:b_phi_delta_m}), the expression in Eq.~(\ref{eq:KaiserBias}) is \textit{not} symmetric in $(\delta_m)_R-\mu$. Consequently, $b_1$ changes when we split the sample into two equal-sized subpopulations. In fact, since $\Delta b_1 < 0$ for halos in the subpopulation with $(\delta_m)_R < \mu$ and $\Delta b_1 > 0$ for halos in the complementary subpopulation, the difference between the two $b_1$ values can be sizable. 

To validate our results, we ran several $N$-body simulations to explicitly compute the derivative in Eq.~(\ref{eq:SUb_phi}) for the different subpopulations and also measured how $b_1$ varies across halos in different dark matter environments. We discuss these simulations in more detail in the following section. The results, shown in Fig.~\ref{fig:DarkMatterSplit}, are in remarkable agreement with the theoretical prediction and show that $b_\phi$ depends quadratically on the large scale dark matter environment, whereas $b_1$ depends on it linearly.  This difference in dependence of these bias parameters on the environment is precisely what enables strong multitracer gains for our method.


\section{Validation with $N$-Body Simulations}\label{sec:ValSim}

\subsection{Robustness of the Dark Matter Split}
To validate the theoretical claims outlined in Sec.~\ref{sec:TheoBack}, we performed a set of $N$-body simulations. Motivated by recent results in the literature \cite{Merino_2026}, which suggested that the separate-universe derivative of Eq.~\eqref{eq:SUb_phi} and a direct power-spectrum inference of $b_\phi$ may differ for environmentally selected samples, we compute the environment dependence of $b_\phi$ in several independent ways.

First, we compute the derivative in Eq.~\eqref{eq:SUb_phi} using a standard finite-difference scheme,
\begin{equation}
    b_\phi
    = 2 \frac{\partial \log \bar n}{\partial \log \sigma_8}
    \approx
    2\left\langle
    \frac{ \sigma_8}{\bar n}\,
    \frac{\bar n^{+}-\bar n^{-}}{ \sigma_8^{+}- \sigma_8^{-}}
    \right\rangle_{\rm sims},
    \label{eq:FiniteDiff}
\end{equation}
where the average is taken over many initial-condition seeds and $\bar n = (\bar n^+ + \bar n^-)/2$. Here, $\bar n^{\pm}$ denotes the number density of the tracer measured in simulations with
$\sigma_8 \rightarrow \sigma_8(1\pm \Delta)$, while all other cosmological parameters are held fixed. To reduce cosmic variance, we compute $\bar n^{+}$ and $\bar n^{-}$ from simulations that share the same initial-condition seeds.

We ran 20 (\textit{i.e.} 10 seeds with two paired realisations) \texttt{FastPM} simulations in boxes of volume $(512\,\mathrm{Mpc}/h)^3$, using a mesh of $256^3$ cells. This allows us to accurately resolve halos with masses down to
$\sim 1.4\cdot 10^{13}\,M_\odot/h$ (the $20$-particle \texttt{FoF} floor). We evolved the dark matter to redshift $z=0.5$ and used a Friends-of-Friends (\texttt{FoF}) halo finder (as implemented in \texttt{nbodykit} \cite{Hand_2018}) with linking length $\ell=0.2$ to identify halos and estimate their masses. We evaluate Eq.~\eqref{eq:FiniteDiff} around the fiducial \texttt{Planck2015} cosmology with $\Delta=0.05$.
Defining the tracer requires some care. If we were to split halos into quantile bins separately in each cosmology, the finite-difference estimator would revert to the $b_\phi$ of the total sample plus additional noise. To avoid this, we use {fixed bins} across all simulations, so that the tracer population is defined consistently and self-containedly when $\sigma_8$ is varied. This isolates the response of the \emph{same} selected population under changing $\sigma_8$.

\begin{figure}
    \includegraphics[width=1\linewidth]{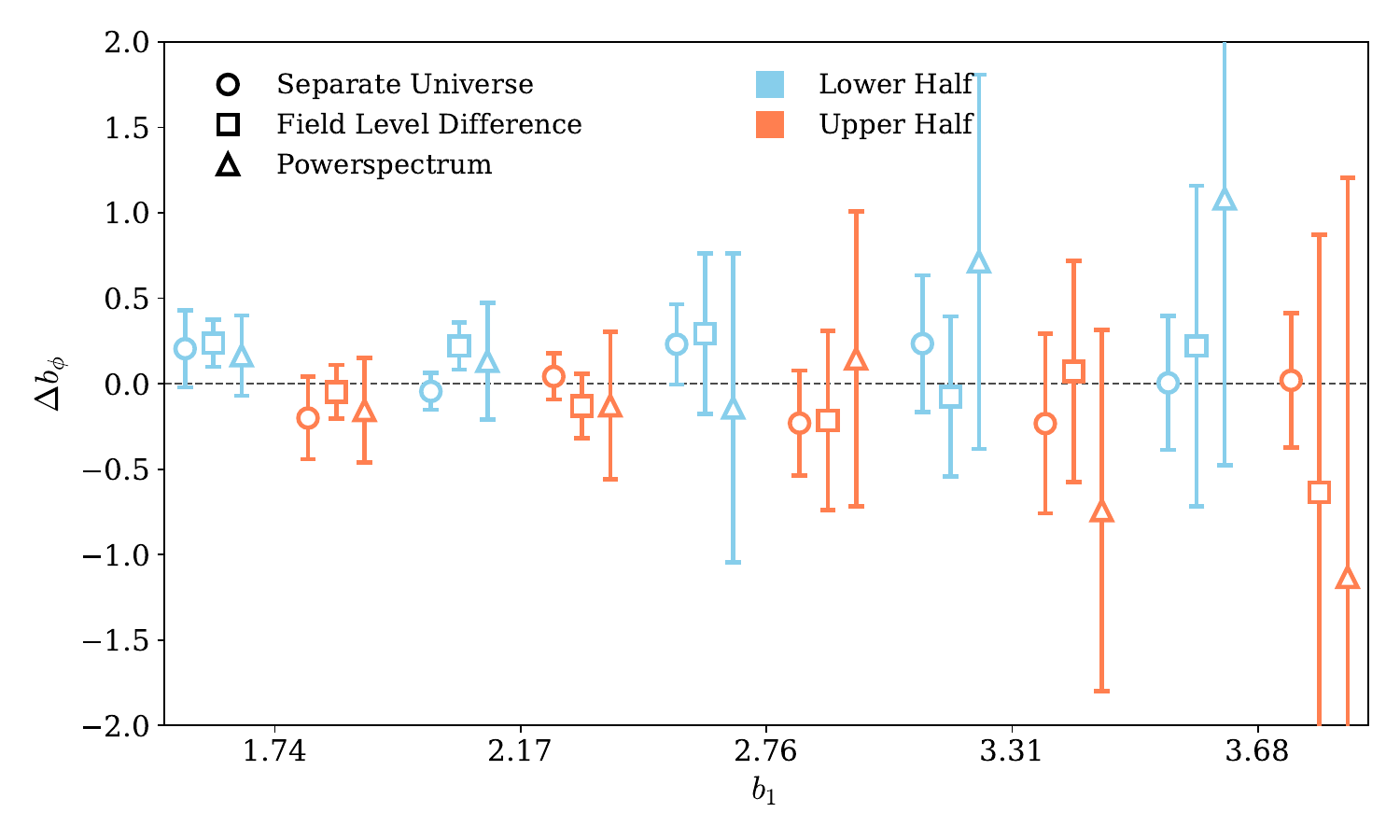}
    \caption{Validation of the claim that $b_\phi$ does not change when a tracer population is split into two equal-sized subpopulations by its large-scale dark matter environment. We show the difference $\Delta b_\phi$ between the $\fnl$ response of the \textit{high} ($(\delta_m)_R>\mu$, coral) and \textit{low} ($(\delta_m)_R\leq\mu$, skyblue) subpopulations and the $\fnl$ response of the total sample, as a function of linear bias. The five linear bias values correspond to halos in five distinct mass bins chosen consistently between the \texttt{FastPM} and \texttt{Quijote} simulations. All inference methods (directly from the scale-dependent bias, shown by the square and triangular data points, and indirectly from the finite-difference derivative, shown by the circular data points) agree with $\Delta b_\phi=0$ for all mass bins. Error bars are estimated from the scatter across the 10 independent simulation seeds.}
    \label{fig:RobustnessCheck}
\end{figure}

Second, to infer $b_\phi$ from the power spectrum we use 20 (\textit{i.e.} 10 seeds with paired values of $\fnl$) \texttt{Quijote-PNG} simulations run in boxes of volume $(1\,\mathrm{Gpc}/h)^3$ with $\fnl=\pm 100$ \cite{Villaescusa_Navarro_2020, Pylians, Coulton_2023}. We estimate $b_\phi$ in two complementary ways. In the first approach, we measure the cross-spectrum
\begin{equation}
    P_{tm}^{\pm}(k)
    =
    \left(b_1 \pm b_\phi \frac{100}{\mathcal M(k)}\right) P_{mm}(k),
\end{equation}
and then infer $b_\phi$ with the optimal estimator~\eqref{eq:Estimator2} from Appendix~\ref{app:OptEst}, fitting over scales $k<0.03\,h/\mathrm{Mpc}$. To reduce residual dependence on $\fnl$ and to improve the signal-to-noise ratio, we average the inferred $b_\phi$ values from the $\fnl=\pm 100$ runs and over the 10 paired seeds.

In the second approach, we construct a field-level difference between the tracer overdensity fields measured in the two simulations,
$\delta_t(\fnl=100)$ and $\delta_t(\fnl=-100)$. The shared linear-bias contribution $b_1\delta_m$ cancels, leaving
\begin{equation}
    \Delta(k)
    \equiv
    \delta_t(\fnl=100,k)-\delta_t(\fnl=-100,k)
    =
    200\,b_\phi\,\phi(k)+\varepsilon + \mathcal O(\fnl^2).
\end{equation}
By measuring the cross-correlation of $\Delta(k)$ with the shared dark matter field, we infer $b_\phi$ using the corresponding optimal estimator~\eqref{eq:Estimator3} from Appendix~\ref{app:OptEst}, again fitting over ${k<0.03\,h/\mathrm{Mpc}}$. We average over the 10 seeds to reduce scatter.

The simulation results are summarized in Fig.~\ref{fig:RobustnessCheck}. We find strong agreement between all methods and a robust validation of the theoretical prediction that $b_\phi$ does not change when the tracer population is split into two equal-sized subpopulations by its large-scale dark matter environment. We verify this statement across a range of halo mass bins selected consistently between the \texttt{Quijote} and \texttt{FastPM} runs, while ensuring that the linear bias of the halo subpopulations matches between the two simulation suites.

\subsection{On Splitting by the Halo Density}
A natural question is whether one could obtain similar multitracer benefits by splitting halos based on their \textit{own} smoothed overdensity field $(\delta_h)_R$. 
Since, on the large scales relevant for the $\fnl$ signal, $(\delta_h)_R$ and $(\delta_m)_R$ are related simply by the (scale-independent) linear bias when $\fnl=0$, one might indeed expect the two approaches to yield nearly equivalent results. 
Splitting by $(\delta_h)_R$ would be especially appealing observationally, and this idea was also explored in \cite{Castorina_2018}. 
In the following, we show why this strategy does not succeed in practice and provide an intuitive explanation for this rather counter-intuitive outcome.

\begin{figure}
    \centering
    \includegraphics[width=0.49\linewidth]{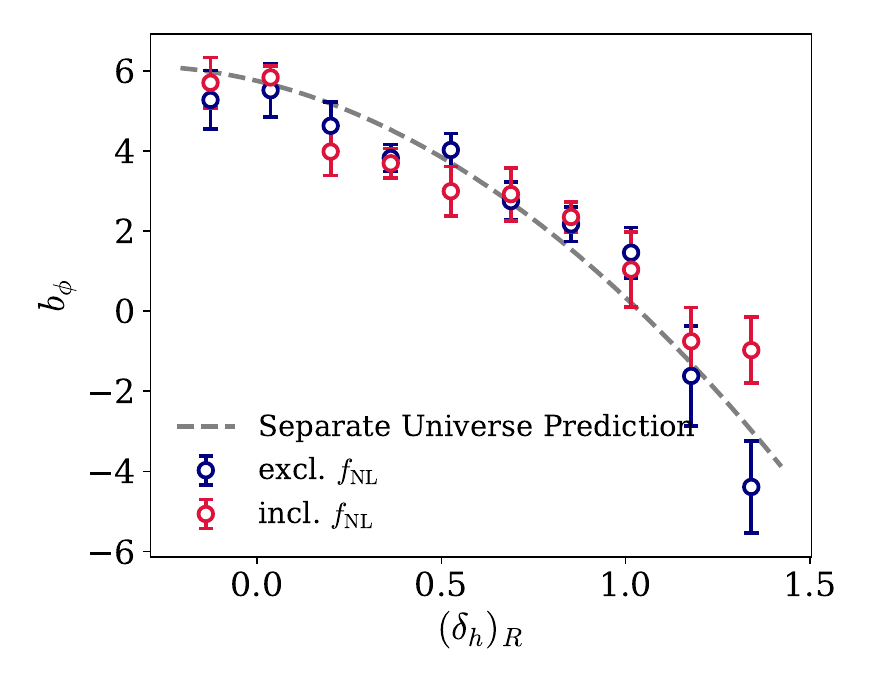}
    \hfill
    \includegraphics[width=0.49\linewidth]{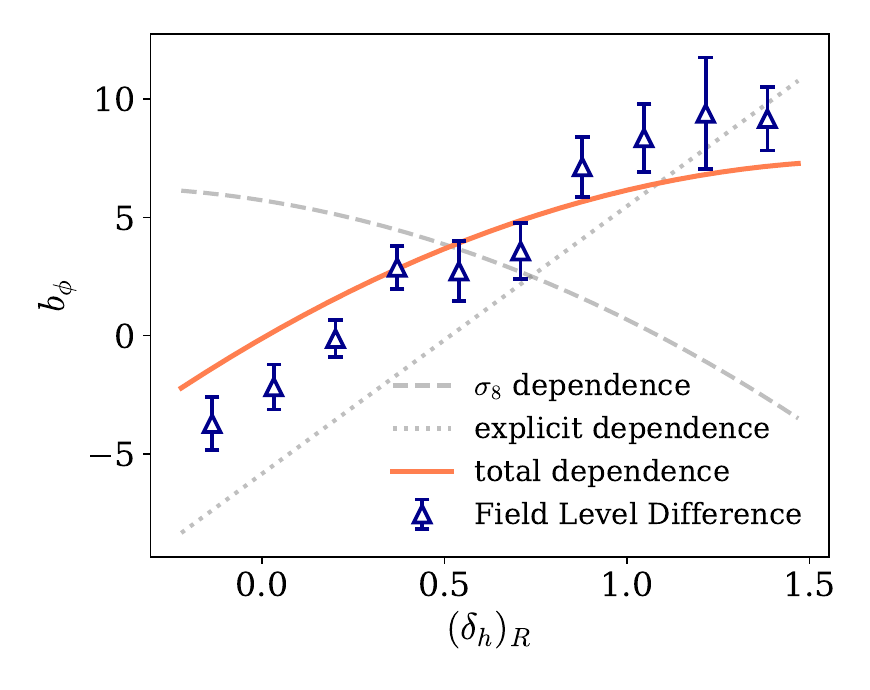}
    \caption{Dependence of the $\fnl$ response on the large-scale halo environment $(\delta_h)_R$. 
    On the left we show results of the finite-difference computation~\eqref{eq:FiniteDiff} for halos in the mass bin
    ${M \in [1.5\cdot 10^{13}, 4.64\cdot 10^{14}]\,M_\odot/h}$ at redshift $z=0.5$, selected by their own overdensity field. We show this dependence both in simulations run without (blue dots) and with (red dots) PNG to illustrate that these two tracer definitions have only a small effect on $b_\phi$. 
    To compute the finite-difference derivatives we ran an additional set of 20 \texttt{FastPM} simulations, in which we injected $\fnl=100$.
    On the right we show results of inferring $b_\phi$ directly from the \texttt{Quijote-PNG} simulations, using the estimator~\eqref{eq:Estimator3}. 
    The disagreement between the two approaches is explained in the main text.
    For these plots we smoothed the halo overdensity field with $R=20\,\mathrm{Mpc}/h$ to reduce scatter and obtain agreement with the linear-theory prediction.}
    \label{fig:DepHaloEnv}
\end{figure}

We begin by noting that the halo environment selection can be treated analogously to the matter-environment analysis in Sec.~\ref{sec:TheoBack}. 
Consider the subpopulation of halos for which the large-scale environment takes the value $\delta$, and denote by $p_h(\delta)$ the probability that $(\delta_h)_R=\delta$. 
The number density of the selected subpopulation is then $\bar n_{\rm sel}=p_h(\delta)\,\bar n$. 
Using Eq.~\eqref{eq:SUb_phi}, this implies
\begin{equation}
    \Delta b_\phi
    =
    2\frac{\partial \log p_h}{\partial \log \sigma_8}.
    \label{eq:SU_bphi_dh}
\end{equation}
Inferring the conditional distribution of halos from the bias expansion
$(\delta_h)_R = b_1(\delta_m)_R + (\varepsilon)_R$
we can write, as in the previous section,
\begin{equation}
    (\delta_h)_R\,\big|\, \text{halo formed} \sim \mathcal N(\mu_h,\sigma_h^2),
\end{equation}
where $\mu_h$ and $\sigma_h$ depend on the conditional distribution of the dark matter at halo positions, the linear bias, and the shot noise. 
Computing Eq.~\eqref{eq:SU_bphi_dh} therefore requires accounting for the $\sigma_8$-dependence of both the shot noise and the linear bias, which makes the calculation considerably more involved than for the matter-split case. 
The details are given in Appendix~\ref{app:b_phi_of_dh}. 
We compare the resulting separate-universe prediction for $b_\phi[(\delta_h)_R]$ to finite-difference derivatives and find remarkable agreement, as shown in the left panel of Fig.~\ref{fig:DepHaloEnv}. 
This agreement holds whether we consider simulations with $\fnl=0$ or $\fnl=100$.

However, when we compare the separate-universe prediction for $b_\phi$ as a function of $(\delta_h)_R$ to the dependence extracted from power-spectrum fits in the \texttt{Quijote-PNG} simulations, we find a strong disagreement (right panel of Fig.~\ref{fig:DepHaloEnv}). 
A similar discrepancy was already noted in the literature \cite{Merino_2026, morawetz2024constrainingprimordialnongaussianitydensitysplit}. 
Within the present formalism, the origin of the inconsistency can be traced to an \emph{explicit} dependence of the conditional mean of the halo-environment PDF on the long-wavelength potential.

Specifically, because the mean of the conditional distribution $p_h$ depends explicitly on the gravitational potential through
\begin{equation}
    \mu_h \supseteq b_\phi\,\fnl\,\phi_\ell,
\end{equation}
the derivative in Eq.~\eqref{eq:SU_bphi_dh} is modified. 
The $\fnl$ response of the selected abundance becomes
\begin{equation}
    \Delta b_\phi
    =
    \frac{d \log p_h}{d(\fnl \phi_\ell)}
    =
    \frac{\partial \log p_h}{\partial(\fnl \phi_\ell)}
    + 2\frac{\partial \log p_h}{\partial \log \sigma_8}.
\end{equation}
The second term is the usual contribution that would be present if the $\fnl$ dependence entered only through $\sigma_8$. 
However, the first term is non-vanishing due to the explicit shift of $\mu_h$. 
It can be computed from
\begin{equation}
    \frac{\partial \log p_h}{\partial(\fnl \phi_\ell)}
    =
    \frac{\partial \mu_h}{\partial(\fnl \phi_\ell)}
    \frac{\partial \log p_h}{\partial \mu_h}\,
    =
    b_\phi\,
    \frac{(\delta_h)_R-\mu_h}{\sigma_h^2},
    \label{eq:ExplPiece}
\end{equation}
where we used $\partial \mu_h/\partial(\fnl \phi_\ell)=b_\phi$.
Only after including this additional term, linear in $(\delta_h)_R-\mu_h$, can the separate-universe prediction reproduce the dependence inferred from the \texttt{Quijote-PNG} simulations, as shown in Fig.~\ref{fig:DepHaloEnv}. We note that the halo field is substantially more non-Gaussian than the late-time dark matter field, so the agreement between our Gaussian description and simulations is necessarily more qualitative than in the previous section. Nevertheless, our central claim, that the dependence of $b_\phi$ on $\delta_h$ is not well captured by the $\sigma_8$ derivative, is robust and reproduced even within this approximation.

The key consequence of Eq.~\eqref{eq:ExplPiece} is that the explicit piece forces the resulting dependence of $b_\phi$ on $(\delta_h)_R$ to be roughly linear. 
Since, as shown in the previous section, the linear bias $b_1$ also depends linearly on $(\delta_h)_R$, it follows that the ratio remains approximately constant for any selected subpopulation:
\begin{equation}
    \frac{b_1^{\rm sel}}{b_\phi^{\rm sel}}
    \approx
    \frac{b_1^{\rm tot}}{b_\phi^{\rm tot}}.
\end{equation}
This near-constancy of $b_1^{\rm sel}/b_\phi^{\rm sel}$ for halos selected by their own halo environment was already observed in numerical studies \cite{morawetz2024constrainingprimordialnongaussianitydensitysplit} and for zero-bias tracers in \cite{Merino_2026}.
Specifically, since $b_1^{\rm tot}/b_\phi^{\rm tot}$ is finite except 
at $b_1^{\rm tot}=1$, $b_\phi^{\rm tot}=0$, we find that for  $b_1^{\rm sel}=0$ one must have $b_\phi^{\rm sel} \sim 0$, in agreement with \cite{Merino_2026}. We note that earlier 
analysis in \cite{Castorina_2018} that found $b_\phi \ne 0$ for zero bias tracers defined by halo overdensity used a separate universe 
approximation to determine $b_\phi$, 
which does not fully capture the 
$f_{\mathrm{NL}}$ response (Fig. \ref{fig:DepHaloEnv}). 

Multitracer analyses improve upon single-tracer constraints only when the ratio $b_1/b_\phi$ varies across the different tracers. 
The approximately constant ratio derived above therefore implies that splitting by the halo environment cannot yield multitracer gains. 
To build intuition for this rather counter-intuitive result, we make two remarks. 
First, on the large scales where the $\fnl$ signal is strongest, the halo field $\delta_h$ is well approximated as a Gaussian field \cite{Bernardeau_2002}. 
For Gaussian fields, the power spectrum is a lossless summary statistic \cite{Tegmark_1997}. 
Therefore, if one only considers the large-scale halo field, it is impossible to improve beyond the single-tracer constraints derived from the corresponding power spectrum. 
Second, more generally, for any Gaussian field the overdensity field of a subpopulation selected by the parent field’s overdensity reduces (at the level of correlation functions) to a rescaled version of the original field. 
We demonstrate this explicitly in Appendix~\ref{app:GaussCond}. 
This immediately implies that $b_1^{\rm sel}/b_\phi^{\rm sel}\approx b_1^{\rm tot}/b_\phi^{\rm tot}$ and, moreover, that such rescaled subpopulations do not carry additional information beyond what is already contained in any one tracer.

\section{Observational Considerations}\label{sec:Obs}

Having established that splitting a tracer population by its large-scale dark matter environment can substantially improve constraints on
$f_{\mathrm{NL}}$, we now address observational considerations.
First, we show that because $\delta_h$ and $\delta_m$ are strongly correlated on large scales, a simple transfer-function estimate of $\delta_m$ yields unbiased constraints when used for the environmental split.
We then examine the dependence of the improvement on linear bias and number density and find that achieving the multitracer regime requires roughly $\bar n \sim 10^{-4}\,(h/\mathrm{Mpc})^{3}$, where cosmic variance dominates over shot noise and sample-variance cancellation becomes effective.
Finally, we forecast the expected performance for the DESI DR2 LRG sample, finding that one can reach
$\sigma(f_{\mathrm{NL}})\sim 3$--$4$ with our method, up to a factor of $2$--$3$ improvement over the expected single tracer constraints.

\begin{figure}
    \centering
    \includegraphics[width=0.49\linewidth]{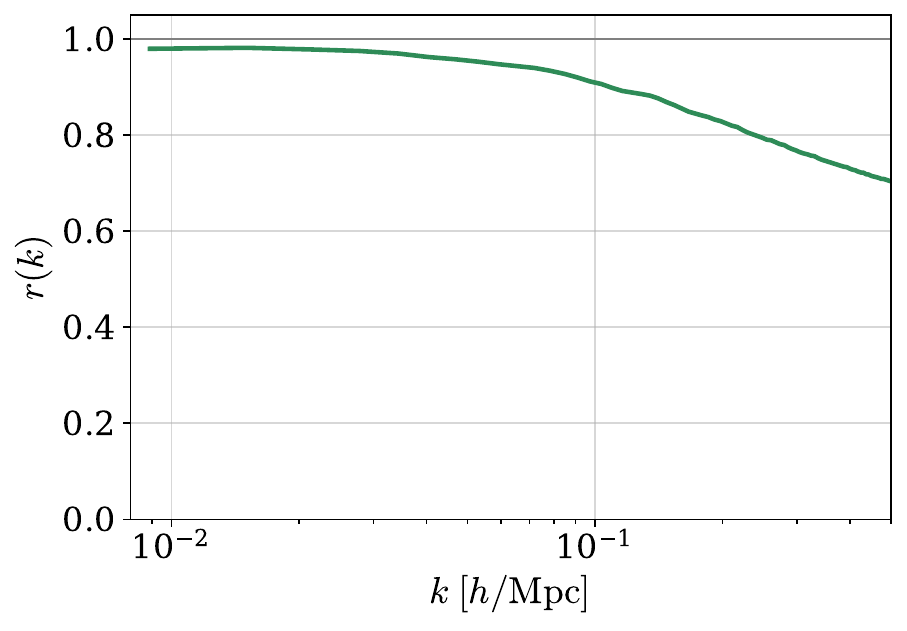}
    \hfill
    \includegraphics[width=0.49\linewidth]{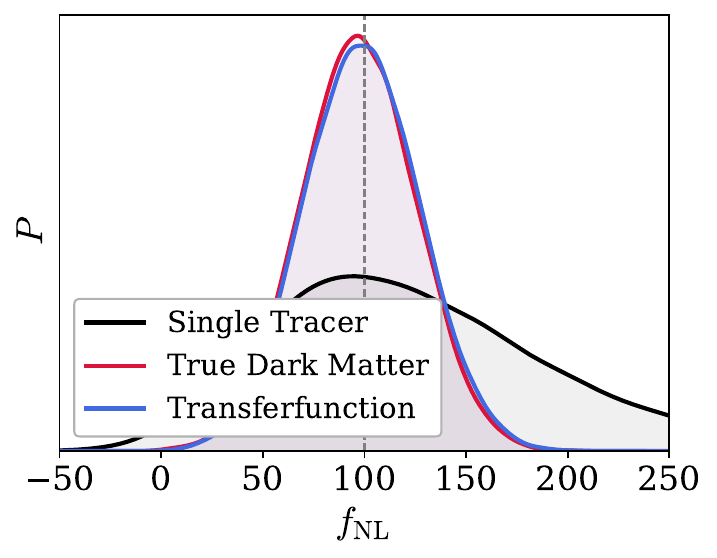}
    \caption{Validation of the linear estimator~\eqref{eq:Tranf} as a basis for the environmental split.
    The \textit{left} panel shows that even in the presence of PNG ($\fnl = 100$) the halo field $\delta_h$ is highly correlated with the underlying matter field $\delta_m$ on large scales, with the cross-correlation coefficient
    ${r(k)=\langle \delta_h^*(k)\delta_m(k)\rangle' /\sqrt{P_{mm}P_{hh}}}$ close to unity until $k\simeq 1/R=0.1\,h/\mathrm{Mpc}$. 
    The \textit{right} panel shows the posterior constraints on $f_{\mathrm{NL}}$ from single- and multitracer fits, using either the true or the inferred dark matter distribution.
    In all cases, $f_{\mathrm{NL}}$ is recovered without bias, and splitting by dark matter (true or inferred) improves the constraints by a factor of $\sim 2.7$.
    Details of the fitting procedure are given in Appendix~\ref{app:Fitting}.}
    \label{fig:TransferPlots}
\end{figure}

Because splitting by the tracer's \emph{own} overdensity cannot improve the constraints (as discussed in Sec.~\ref{sec:ValSim} and Fig.~\ref{fig:DepHaloEnv}), we need to infer the underlying dark matter environment from the halo field.
On the relevant large scales, linear theory provides a reliable description of biased tracers, implying that $\delta_h$ and $\delta_m$ are highly correlated.
This is illustrated in the left panel of Fig.~\ref{fig:TransferPlots}.
In principle, $\delta_m$ can be reconstructed from $\delta_h$ using the minimum-variance estimator
\begin{equation}
    \hat\delta_m(k)=\frac{P_{hm}(k)}{P_{hh}(k)}\,\delta_h(k)
    =\frac{b(k)P_{mm}(k)}{b(k)^2P_{mm}(k)+1/\bar n}\,\delta_h(k),
\end{equation}
where $P_{hm}=b(k)P_{mm}$.
However, in our case the scale-dependent bias
$b(k)=b_1+f_{\mathrm{NL}}\,b_\phi/\mathcal M(k)$ depends on $f_{\mathrm{NL}}$,
so using this estimator risks introducing self-referential dependence on the parameter we aim to constrain.

Instead, we use a simple estimator that does not require choosing a fiducial $f_{\mathrm{NL}}$:
\begin{equation}
    \hat \delta_m(k) = T(k)\,(\delta_h)_R(k), \qquad
    T(k) = \sqrt{\frac{P_{mm}(k)}{P_{hh}(k)}} .
    \label{eq:Tranf}
\end{equation}
Here, $P_{hh}(k)$ is the halo power spectrum taken directly from the data. This definition ensures that, on average, the $f_{\mathrm{NL}}$ dependence is removed from the field used to split the halos in an entirely data-driven manner. This avoids both the circularity of assuming a fiducial $f_{\mathrm{NL}}$ value to perform the split, and the degeneracy encountered in Sec.~\ref{sec:ValSim} when attempting to split halos by their own overdensity field. While this reconstruction method requires substantial refinement to account for redshift-space distortions (RSD) and non-trivial survey windows, our present method is not intended to be an optimal or completely realistic reconstruction. Rather, it serves as a demonstration that a straightforward, linear-theory approach can already yield unbiased constraints.

The transfer-function normalization suppresses the direct $f_{\rm NL}$ dependence of the reconstructed environmental variable, allowing the resulting samples to approximate a split by the underlying matter environment. Because the reconstruction and split are deterministic functions of the halo catalogue, they cannot create information beyond that in the full catalogue. Their gain relative to the conventional single-tracer power spectrum should therefore be interpreted as a nonlinear compression of information not captured by that statistic, rather than as information added by the reconstruction itself. We note that splitting the halo population into two subpopulations is a highly non-linear operation. In the exactly Gaussian continuous-field limit, no net gain would be recovered from such a deterministic transformation. 

For a realistic survey, an optimal reconstruction would require a Gaussian field-level analysis, which can be achieved using the methods presented in \cite{Horowitz_2019, Seljak_2017, Seljak_1998, Millea_2022}. The galaxy field can be modeled as $\boldsymbol \delta_g = \mathbf R \boldsymbol \delta_m + \boldsymbol \epsilon$, where $\mathbf R$ is a linear operator that accounts for RSD, bias, and the survey window, and $\boldsymbol \epsilon$ is a Gaussian noise field. Using a field-level likelihood, the maximum a posteriori (MAP) estimate $\hat{\boldsymbol \delta}_m$ can be found even if the underlying covariance is dense and numerically challenging to invert. This MAP estimate inherently provides a suitably smoothed field that can be used directly to split the halos. Finally, we note that while RSD complicates the analysis, it may also improve the reconstruction quality since peculiar velocities directly trace the dark matter field. We will explore these aspects in future work.

To test this procedure, we perform a multitracer fit on the \texttt{Quijote-PNG} simulations averaged over the 10 runs with $f_{\mathrm{NL}}=100$.
We infer $f_{\mathrm{NL}}$ from a halo population treated as a single tracer, and we construct the environmental split in two ways: using the true $(\delta_m)_R$ and using the inferred $\hat\delta_m$ from Eq.~\eqref{eq:Tranf}.
The results (right panel of Fig.~\ref{fig:TransferPlots}) show that $f_{\mathrm{NL}}$ is recovered without bias in all cases.
The fitting details are provided in Appendix~\ref{app:Fitting}.

Since the estimator~\eqref{eq:Tranf} can be applied directly to data, we now forecast the achievable constraints for tracers that are well approximated as mass-selected halos, which is expected to be a reasonable description of the DESI LRG population \cite{yuan2023desionepercentsurveyexploring}.
To match the DESI LRG effective response as closely as possible, we choose the redshift $z_{\rm eff}=0.74$.
At this redshift, the LRG number density is $\bar n \sim 4\cdot 10^{-4}\,(h/\mathrm{Mpc})^3$ and the linear bias is $b_1\sim 2$ \cite{chaussidon2025constrainingprimordialnongaussianitydesi}; for the survey volume we adopt the projected DESI DR2 LRG value $V\sim 11\,(\mathrm{Gpc}/h)^3$.
Assuming systematics can be controlled, we use modes in the range
\begin{equation}
    k \in [0.003, 0.08]\,h/\mathrm{Mpc},
\end{equation}
where $k_{\max}$ is chosen consistently with the DR1 analysis \cite{chaussidon2025constrainingprimordialnongaussianitydesi}.

We forecast the constraining power on $f_{\mathrm{NL}}$ using a standard Fisher matrix approach.
We consider forecasts with $N\in\{1,\dots,4\}$ tracers, whose covariance is modeled as
\begin{equation}
    C_{ij}(k) = b_i(k)b_j(k)\,P_{mm}(k) + \frac{\delta_{ij}}{\bar n_i},
    \qquad
    b_i(k)=b_1^{(i)} + f_{\mathrm{NL}}\,\frac{b_\phi^{(i)}}{\mathcal M(k)}.
\end{equation}
The Fisher matrix is given by \cite{Tegmark_1997}
\begin{equation}
    F_{\theta_a\theta_b} = \sum_k N_k\,
    \mathrm{tr}\!\left[
    \mathbf C(k)^{-1}\frac{\partial \mathbf C}{\partial \theta_a}\mathbf C(k)^{-1}\frac{\partial \mathbf C}{\partial \theta_b}
    \right],
    \qquad
    N_k = V\frac{k^2\Delta k}{4\pi^2},
\end{equation}
where we take $\boldsymbol\theta=(f_{\mathrm{NL}},b_1^{(1)},\dots,b_1^{(N)})$ and
\begin{align}
    \frac{\partial C_{jl}}{\partial b_1^{(i)}}(k)
    &= \big(\delta_{ji}\,b_l(k) + b_j(k)\,\delta_{li}\big)\,P_{mm}(k),\\
    \frac{\partial C_{ij}}{\partial f_{\mathrm{NL}}}(k)
    &= \left(\frac{b_\phi^{(i)}}{\mathcal M(k)}\,b_j(k) + b_i(k)\,\frac{b_\phi^{(j)}}{\mathcal M(k)}\right)P_{mm}(k).
\end{align}
We evaluate all quantities on a linear $k$ grid with $\Delta k=0.001\,h/\mathrm{Mpc}$ and compute $P_{mm}(k)$ using the linear power spectrum from \texttt{camb} \cite{Lewis_2000}.
The marginalized uncertainty is then
\begin{equation}
    \sigma(f_{\mathrm{NL}})^2 = \left(\mathbf F^{-1}\right)_{f_{\mathrm{NL}}f_{\mathrm{NL}}}.
\end{equation}

\begin{figure}[h!]
    \centering
    \includegraphics[width=1\linewidth]{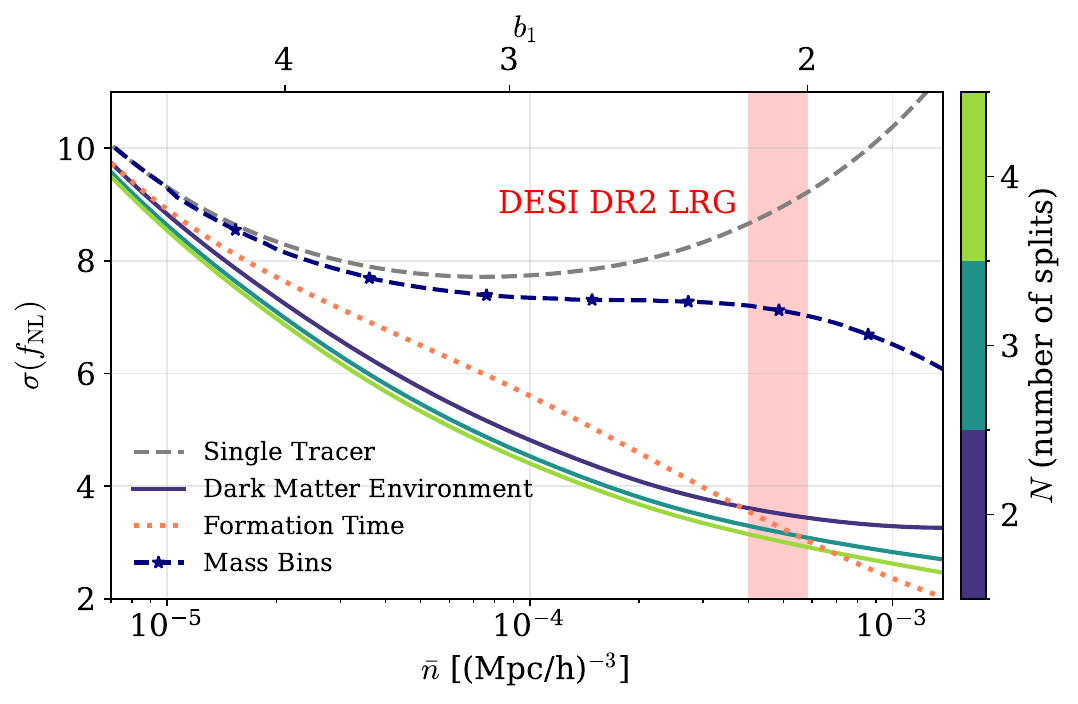}
    \caption{Fisher forecast of the expected improvements from a dark matter environmental split.
    We forecast constraints on $f_{\mathrm{NL}}$ using mass-selected halos at the DESI LRG effective redshift ${z=0.74}$.
    All results are based on the \texttt{Halfdome} simulations.
    The grey dashed curve shows the single-tracer constraints, while the blue dashed line shows the multitracer constraints obtainable from mass binning.
    The orange dotted line shows the constraints obtainable from splitting based on a formation-time dependent selection, where the $b_\phi$ dependence on formation is modeled following \cite{Fondi_2024} and $b_1$ is assumed independent of formation time.
    The solid curves correspond to splits depending on the large-scale dark matter environment.
    The improvement nearly saturates already for a split into two samples, which motivates the focus of this work.
    The DESI LRG sample is indicated by the coral band, with the lower limit chosen to match abundance and the upper limit chosen to match bias. At this number density splitting by the large-scale dark matter environment can substantially improve constraints on $f_{\mathrm{NL}}$, leading to a factor of $2$--$3$ improvement. At higher number densities still, formation time is more informative.\protect\footnotemark }
    \label{fig:Fisher}
\end{figure}

For mass-binned halos, we take $\bar n$ and $b_1$ (and the change in these quantities when splitting by the dark matter environment) directly from $N$-body simulations.
To cover sufficiently high number densities, we use the \texttt{Halfdome} simulations, which resolve halos down to the necessary ${M_{\min}\sim 2\cdot 10^{12}\,M_\odot/h}$ \cite{Bayer_2025}.
For the $b_\phi$--$b_1$ relation, we assume the universality prediction 
\begin{equation}
    b_\phi(b_1)=2\delta_c\,(b_1-p),\qquad p=1,
\end{equation}
with $\delta_c=1.686$.
For the multitracer environmental split, we assume halos are divided into equal-probability bins.
In this case the dependence of the $\fnl$ response on $(\delta_m)_R$ is captured by the normalized Gaussian variable
$\nu \equiv \big((\delta_m)_R-\mu\big)/\sigma$, where the bin edges are determined by Gaussian statistics.
The resulting shift in $b_\phi$ is
\begin{equation}
    \Delta b_\phi =
    \sqrt{\frac{8}{\pi}}\,
    \frac{\nu_1 e^{-\nu_1^2/2}-\nu_0 e^{-\nu_0^2/2}}
    {\mathrm{erfc}(\nu_1/\sqrt 2)-\mathrm{erfc}(\nu_0/\sqrt 2)} ,
\end{equation}\footnotetext{This saturation of information may be understood as follows. When splitting $\delta_h$ by the dark matter environment $\delta_m$, we are probing some of the information contained in the cross spectrum $P_{hm}$. A simple computation shows that the $\fnl$ Fisher information of the $(P_{hh}, P_{hm})$ data vector also saturates at ${\bar n \sim 10^{-3}\,(\mathrm{Mpc}/h)^{-3}}$ at the given redshift. On the contrary, the additional information contained in a split based on formation time is from assumed knowledge of secondary halo properties and therefore need not saturate at the linear theory information floor set by $(P_{hh}, P_{hm})$.}
consistent with the earlier theoretical motivation and validated in simulations (now written for finite bins).
Since $\mu$ and $\sigma$ enter only through the $\nu_i$ (which are fixed by the equal-bin condition), the forecast is insensitive to the particular values of $\mu$ and $\sigma$.
When splitting into $N$ equal-probability bins, the number density scales as $\bar n\rightarrow \bar n/N$ for each tracer.
The resulting forecasts are shown in Fig.~\ref{fig:Fisher}.

A few trends are worth highlighting.
First, in the single-tracer analysis the error increases at high number density.
Although this can seem counterintuitive, it follows from the fact that for mass-selected halos, increasing $\bar n$ drives $b_1\rightarrow 1$, and the universality relation then implies $b_\phi\rightarrow 0$.
Therefore, unless multitracer information is used, a very high number density sample can be less informative for $f_{\mathrm{NL}}$ than a lower number density sample with larger $b_\phi$. This highlights the need for multitracer analyses to optimally extract the $\fnl$ signal in upcoming surveys, which can reach very high number densities. 
Second, multitracer improvements become significant only once the number density reaches roughly
$\bar n \sim 10^{-4}\,(h/\mathrm{Mpc})^{3}$.
At the LRG number densities and biases, the environmental split improves constraints by a factor of $2$--$3$, substantially outperforming simple mass-binning multitracer approaches. We present a more detailed discussion, also analyzing improvements possible for the DESI QSO sample and by combining formation time and density binning, in App.~\ref{app:FurtherFisher}.

\section{Discussion}\label{sec:Conc}
Currently, the most stringent constraints on primordial non-Gaussianity (PNG) from large-scale structure (LSS) are obtained via analyses of the power spectrum of single tracers \cite{chaussidon2025constrainingprimordialnongaussianitydesi}. Such analyses are suboptimal in two ways. First, the galaxy bispectrum can contribute an additional ${\sim}10$--$20\%$ to the constraint of the DESI DR1 data \cite{chudaykin2025reanalyzingdesidr13}. Further improvements can potentially be obtained by extending the analysis to the field-level~\cite{andrews2026fieldlevelinferenceprimordialnongaussianity, chen2026primordialnongaussianityfieldlevelcramerrao, sullivan2025localprimordialnongaussianbias}. Second, single-tracer bias models are inherently limited because they necessarily average over the intrinsic variation of bias parameters across the sample. This can only be improved in the context of multitracer analyses, where most existing proposals rely on secondary halo properties, such as concentration, formation time, or size, to outperform single-tracer results.

In this work we developed and validated an environmental multitracer strategy to constrain local PNG. The approach divides a single tracer sample into sub-samples according to its \emph{large-scale dark matter} environment. Splitting a population by its environment has already been explored for zero-bias tracers \cite{Castorina_2018, Merino_2026} and within the density-split formalism \cite{morawetz2024constrainingprimordialnongaussianitydensitysplit}. In Sec.~\ref{sec:TheoBack} we showed that if the split is implemented as two equal-probability bins of the smoothed matter overdensity $(\delta_m)_R$, then each subpopulation has the same PNG response $b_\phi$ as the parent sample. In contrast, the \emph{linear} bias $b_1$ changes substantially across the environment bins. This combination drives the multitracer gain~\cite{Seljak_2009}.

We validated this theoretical mechanism using $N$-body simulations. In Sec.~\ref{sec:ValSim} we demonstrated that the PNG response remains unchanged when the tracer is split into the corresponding two environment subsamples, quantitatively confirming $\Delta b_\phi=0$ (Fig.~\ref{fig:RobustnessCheck}). Importantly, this agreement holds when $b_\phi$ is estimated using independent inference routes: directly from scale-dependent bias in the PNG simulations and indirectly from abundance derivatives computed via separate-universe-like finite differences. The same formalism clarifies why an apparently equivalent procedure does not deliver these gains, thereby offering insight into conclusions reached in earlier environment-split studies \cite{morawetz2024constrainingprimordialnongaussianitydensitysplit, Merino_2026}. Selecting halos by their \emph{own} smoothed overdensity does not reproduce the multitracer behavior, even though the halo and matter densities are tightly correlated at $\fnl=0$. Conditioning on the halos' own field introduces an explicit dependence of the selection on the long-wavelength potential, i.e.\ a direct entry of the PNG signal into the membership of each bin. This contributes a term to the response that is linear in the halo’s own smoothed overdensity. Since $b_1$ is also linear in the same quantity, the ratio $b_1/b_\phi$ remains nearly constant across the sub-samples, and the contrast that powers the cancellation is lost. Selecting on the surrounding dark matter, whose membership carries no such explicit $\fnl$ dependence on linear scales, avoids this effect and preserves the contrast in $b_1$ at fixed $b_\phi$.

To connect with observations, we must construct the environment variable from the tracer field. Sec.~\ref{sec:Obs} addresses this by showing that a simple linear-theory reconstruction of $(\delta_m)_R$, using the transfer-function estimator~\eqref{eq:Tranf}, yields unbiased constraints on $f_{\mathrm{NL}}$. We explicitly tested this in \texttt{Quijote-PNG} simulations. The recovered $f_{\mathrm{NL}}$ is unbiased whether one performs the split using the true $(\delta_m)_R$ or the reconstructed $\hat\delta_m$ (Fig.~\ref{fig:TransferPlots}, right panel). Moreover, the improvement factor is similar in both cases, indicating that nearly the full information gain survives the reconstruction step. 
With this observationally viable split at hand, we forecast the performance for the DESI LRG sample. Using the projected DESI DR2 LRG survey volume and number density at the effective redshift $z_{\rm eff}=0.74$, and restricting to a linear-mode range, our Fisher forecasts show an improvement of roughly a factor of $2$--$3$ in the uncertainty on $f_{\mathrm{NL}}$ compared to single-tracer power-spectrum constraints, reaching $\sigma(f_{\mathrm{NL}})\sim 3$--$4$ in our idealized forecast (Fig.~\ref{fig:Fisher}).

Achieving the forecasted gains is complicated by several observational considerations. First, because the environment selection is constructed from a reconstructed density field, any imperfection in that reconstruction can distort the environmental partition. This includes survey systematics that introduce spurious large-scale correlations. Establishing the robustness of the reconstructed environment under such contaminants is essential. Second, the analysis presented here assumes real-space clustering; a consistent treatment of redshift-space distortions is required. Finally, translating a measured scale-dependent response into a value of $\fnl$ requires knowledge of $b_\phi$, which, as stressed in the introduction, is sensitive to uncertain secondary and astrophysical properties, to assembly bias, and to the $\fnl$ dependence of the halo occupation. This concern is tempered for the present application because LRGs are well approximated as mass-selected halos \cite{yuan2023desionepercentsurveyexploring} and therefore plausibly carry small $b_\phi$ modeling uncertainty. Moreover, the method does not require $b_\phi$ to be known a priori: because $b_\phi$ is the same for both subsamples, the multitracer fit directly constrains the product $b_\phi\,f_{\rm NL}$. A subsequent external prior on $b_\phi$, from simulations or theoretical modeling, then converts this into a constraint on $f_{\rm NL}$ itself.

Unlike multitracer methods that use secondary halo properties, our split is defined using only a linear-theory estimate of the large-scale dark matter environment from the observed tracer field. This avoids modeling assembly bias while still attaining constraints close to those from a formation-time split. Overall, these results suggest that environment-selected multitracer clustering can target the $\sigma(f_{\rm NL})\sim 1$ regime with upcoming galaxy surveys.

\acknowledgments

We would like to thank James Sullivan, Edmond Chaussidon, Richard Feder, Licia Verde, Moritz Münchmeyer and Simone Ferraro for many insightful discussions. This research used resources of the National Energy Research Scientific Computing Center (NERSC), a Department of Energy Office of Science User Facility. We acknowledge the use of the Anthropic LLM \texttt{Claude Opus 4.8} during code implementation. This work is supported by NSF CDSE grant number AST-2408026 and NASA TCAN grant number 80NSSC24K0101.

\bibliographystyle{JHEP}
\bibliography{bibliography}

@misc{planckcollaboration2019planck2018resultsix,
      title="{Planck 2018 results. IX. Constraints on primordial non-Gaussianity}", 
      author="{Planck Collaboration et al.}",
      year={2019},
      eprint={1905.05697},
      archivePrefix={arXiv},
      primaryClass={astro-ph.CO},
      url={https://arxiv.org/abs/1905.05697}, 
}

@article{Sullivan:2023qjr,
    author = "Sullivan, James M. and Prijon, Tijan and Seljak, Uro\v{s}",
    title = "{Learning to Concentrate: Multi-tracer Forecasts on Local Primordial Non-Gaussianity with Machine-Learned Bias}",
    eprint = "2303.08901",
    archivePrefix = "arXiv",
    primaryClass = "astro-ph.CO",
    doi = "10.1088/1475-7516/2023/08/004",
    journal = "JCAP",
    volume = "08",
    pages = "004",
    year = "2023"
}

@article{Fondi_2024,
   title="{Taming assembly bias for primordial non-Gaussianity}",
   volume={2024},
   ISSN={1475-7516},
   url={http://dx.doi.org/10.1088/1475-7516/2024/02/048},
   DOI={10.1088/1475-7516/2024/02/048},
   number={02},
   journal={Journal of Cosmology and Astroparticle Physics},
   publisher={IOP Publishing},
   author={Fondi, Emanuele and Verde, Licia and Villaescusa-Navarro, Francisco and Baldi, Marco and Coulton, William R. and Jung, Gabriel and Karagiannis, Dionysios and Liguori, Michele and Ravenni, Andrea and Wandelt, Benjamin D.},
   year={2024},
   month=Feb, pages={048} }

@article{Voivodic_2021,
   title="{Responses of Halo Occupation Distributions: a new ingredient in the halo model; the impact on galaxy bias}",
   volume={2021},
   ISSN={1475-7516},
   url={http://dx.doi.org/10.1088/1475-7516/2021/05/069},
   DOI={10.1088/1475-7516/2021/05/069},
   number={05},
   journal={Journal of Cosmology and Astroparticle Physics},
   publisher={IOP Publishing},
   author={Voivodic, Rodrigo and Barreira, Alexandre},
   year={2021},
   month=May, pages={069} }

@article{Sullivan:2025fie,
    author = "Sullivan, James M. and Seljak, Uros",
    title = "{Local primordial non-Gaussian bias from time evolution}",
    eprint = "2503.21736",
    archivePrefix = "arXiv",
    primaryClass = "astro-ph.CO",
    doi = "10.1103/8jvj-66zc",
    journal = "Phys. Rev. D",
    volume = "112",
    number = "8",
    pages = "083522",
    year = "2025"
}

@misc{dalal2025estimatingnongaussianbiasusing,
      title="{Estimating non-gaussian bias using counts of tracers}", 
      author={Neal Dalal and Will J. Percival},
      year={2025},
      eprint={2503.21024},
      archivePrefix={arXiv},
      primaryClass={astro-ph.CO},
      url={https://arxiv.org/abs/2503.21024}, 
}

@article{Lazeyras_2023,
   title="{Assembly bias in the local PNG halo bias and its implication for fNL constraints}",
   volume={2023},
   ISSN={1475-7516},
   url={http://dx.doi.org/10.1088/1475-7516/2023/01/023},
   DOI={10.1088/1475-7516/2023/01/023},
   number={01},
   journal={Journal of Cosmology and Astroparticle Physics},
   publisher={IOP Publishing},
   author={Lazeyras, Titouan and Barreira, Alexandre and Schmidt, Fabian and Desjacques, Vincent},
   year={2023},
   month=Jan, pages={023} }

@article{Reid_2010,
   title="{Non-Gaussian halo assembly bias}",
   volume={2010},
   ISSN={1475-7516},
   url={http://dx.doi.org/10.1088/1475-7516/2010/07/013},
   DOI={10.1088/1475-7516/2010/07/013},
   number={07},
   journal={Journal of Cosmology and Astroparticle Physics},
   publisher={IOP Publishing},
   author={Reid, Beth A and Verde, Licia and Dolag, Klaus and Matarrese, Sabino and Moscardini, Lauro},
   year={2010},
   month=Jul, pages={013–013} }

@article{Slosar_2008,
   title="{Constraints on local primordial non-Gaussianity from large scale structure}",
   volume={2008},
   ISSN={1475-7516},
   url={http://dx.doi.org/10.1088/1475-7516/2008/08/031},
   DOI={10.1088/1475-7516/2008/08/031},
   number={08},
   journal={Journal of Cosmology and Astroparticle Physics},
   publisher={IOP Publishing},
   author={Slosar, Anže and Hirata, Christopher and Seljak, Uroš and Ho, Shirley and Padmanabhan, Nikhil},
   year={2008},
   month=Aug, pages={031} }

@article{Seljak_2009,
   title="{Extracting Primordial Non-Gaussianity without Cosmic Variance}",
   volume={102},
   ISSN={1079-7114},
   url={http://dx.doi.org/10.1103/PhysRevLett.102.021302},
   DOI={10.1103/physrevlett.102.021302},
   number={2},
   journal={Physical Review Letters},
   publisher={American Physical Society (APS)},
   author={Seljak, Uroš},
   year={2009},
   month=Jan }

@article{Castorina_2018,
   title="{Primordial Non-Gaussianities and Zero-Bias Tracers of the Large-Scale Structure}",
   volume={121},
   ISSN={1079-7114},
   url={http://dx.doi.org/10.1103/PhysRevLett.121.101301},
   DOI={10.1103/physrevlett.121.101301},
   number={10},
   journal={Physical Review Letters},
   publisher={American Physical Society (APS)},
   author={Castorina, Emanuele and Feng, Yu and Seljak, Uroš and Villaescusa-Navarro, Francisco},
   year={2018},
   month=Sep }

@misc{morawetz2024constrainingprimordialnongaussianitydensitysplit,
      title="{Constraining Primordial Non-Gaussianity with Density-Split Clustering}", 
      author={James Morawetz and Enrique Paillas and Will J. Percival},
      year={2024},
      eprint={2409.13583},
      archivePrefix={arXiv},
      primaryClass={astro-ph.CO},
      url={https://arxiv.org/abs/2409.13583}, 
}

@article{Merino_2026,
   title="{Unclustered tracers remain unclustered: The lack of primordial non-Gaussianity response of bias-zero tracers}",
   volume={707},
   ISSN={1432-0746},
   url={http://dx.doi.org/10.1051/0004-6361/202558825},
   DOI={10.1051/0004-6361/202558825},
   journal={Astronomy \& Astrophysics},
   publisher={EDP Sciences},
   author={Merino, C. and Avila, S. and Adame, A. G. and Anguren, A. and Gonzalez-Perez, V. and Meneses-Rizo, J.},
   year={2026},
   month=Mar, pages={L15} }

@misc{kvasiuk2024talefieldsneuralnetworkenhanced,
      title="{A Tale of Two Fields: Neural Network-Enhanced non-Gaussianity Search with Halos}", 
      author={Yurii Kvasiuk and Moritz Münchmeyer and Kendrick Smith},
      year={2024},
      eprint={2410.01007},
      archivePrefix={arXiv},
      primaryClass={astro-ph.CO},
      url={https://arxiv.org/abs/2410.01007}, 
}

@misc{fondi2026assemblybiaslocalprimordial,
      title="{Assembly bias and local Primordial non-Gaussianity from DESI DR1 Quasars}", 
      author={E. Fondi and L. Verde and E. Chaussidon and J. Aguilar and S. Ahlen and S. BenZvi and D. Bianchi and D. Brooks and T. Claybaugh and A. Cuceu and A. de la Macorra and P. Doel and S. Ferraro and J. E. Forero-Romero and E. Gaztañaga and S. Gontcho A Gontcho and G. Gutierrez and H. K. Herrera-Alcantar and D. Huterer and M. Ishak and R. Joyce and A. Kremin and O. Lahav and C. Lamman and M. Landriau and L. Le Guillou and M. Manera and P. Martini and A. Meisner and R. Miquel and S. Nadathur and N. Palanque-Delabrouille and W. J. Percival and F. Prada and I. Pérez-Ràfols and G. Rossi and L. Samushia and E. Sanchez and D. Schlegel and D. Sprayberry and G. Tarlé and B. A. Weaver and H. Zou},
      year={2026},
      eprint={2602.12357},
      archivePrefix={arXiv},
      primaryClass={astro-ph.CO},
      url={https://arxiv.org/abs/2602.12357}, 
}

@book{daley2008introduction,
  title     = "{An Introduction to the Theory of Point Processes. Volume II: General Theory and Structure}",
  author    = {Daley, D. J. and Vere-Jones, D.},
  year      = {2008},
  publisher = {Springer New York},
  isbn      = {978-0-387-21337-8},
  doi       = {10.1007/978-0-387-49835-5}
}

@misc{SimoPNG,
      title="{Refining local-type primordial non-Gaussianity: Sharpened $b_\phi$ constraints through bias expansion}", 
      author={Boryana Hadzhiyska and Simone Ferraro},
      year={2025},
      eprint={2501.14873},
      archivePrefix={arXiv},
      primaryClass={astro-ph.CO},
      url={https://arxiv.org/abs/2501.14873}, 
}

@article{Desjacques_2018,
   title="{Large-scale galaxy bias}",
   volume={733},
   ISSN={0370-1573},
   url={http://dx.doi.org/10.1016/j.physrep.2017.12.002},
   DOI={10.1016/j.physrep.2017.12.002},
   journal={Physics Reports},
   publisher={Elsevier BV},
   author={Desjacques, Vincent and Jeong, Donghui and Schmidt, Fabian},
   year={2018},
   month=Feb, pages={1–193} }

@article{Akrami_2020,
    author = "Akrami, Y. and others",
    collaboration = "Planck",
    title = "{Planck 2018 results. X. Constraints on inflation}",
    eprint = "1807.06211",
    archivePrefix = "arXiv",
    primaryClass = "astro-ph.CO",
    doi = "10.1051/0004-6361/201833887",
    journal = "Astron. Astrophys.",
    volume = "641",
    pages = "A10",
    year = "2020"
}

@article{Aghanim_2020,
    author = "Aghanim, N. and others",
    collaboration = "Planck",
    title = "{Planck 2018 results. VI. Cosmological parameters}",
    eprint = "1807.06209",
    archivePrefix = "arXiv",
    primaryClass = "astro-ph.CO",
    doi = "10.1051/0004-6361/201833910",
    journal = "Astron. Astrophys.",
    volume = "641",
    pages = "A6",
    year = "2020",
    note = "[Erratum: Astron.Astrophys. 652, C4 (2021)]"
}

@inproceedings{Baumann_2009,
    author = "Baumann, Daniel",
    title = "{TASI Lectures on Inflation}",
    booktitle = "{Theoretical Advanced Study Institute in Elementary Particle Physics: Physics of the Large and the Small}",
    eprint = "0907.5424",
    archivePrefix = "arXiv",
    primaryClass = "hep-th",
    doi = "10.1142/9789814327183_0010",
    pages = "523--686",
    year = "2011"
}

@article{Maldacena_2003,
    author = "Maldacena, Juan Martin",
    title = "{Non-Gaussian features of primordial fluctuations in single field inflationary models}",
    eprint = "astro-ph/0210603",
    archivePrefix = "arXiv",
    doi = "10.1088/1126-6708/2003/05/013",
    journal = "JHEP",
    volume = "05",
    pages = "013",
    year = "2003"
}

@article{Bartolo_2004,
    author = "Bartolo, N. and Komatsu, E. and Matarrese, Sabino and Riotto, A.",
    title = "{Non-Gaussianity from inflation: Theory and observations}",
    eprint = "astro-ph/0406398",
    archivePrefix = "arXiv",
    doi = "10.1016/j.physrep.2004.08.022",
    journal = "Phys. Rept.",
    volume = "402",
    pages = "103--266",
    year = "2004"
}

@article{Komatsu_2001,
    author = "Komatsu, Eiichiro and Spergel, David N.",
    title = "{Acoustic signatures in the primary microwave background bispectrum}",
    eprint = "astro-ph/0005036",
    archivePrefix = "arXiv",
    doi = "10.1103/PhysRevD.63.063002",
    journal = "Phys. Rev. D",
    volume = "63",
    pages = "063002",
    year = "2001"
}

@article{Dalal_2008,
    author = "Dalal, Neal and Dore, Olivier and Huterer, Dragan and Shirokov, Alexander",
    title = "{The imprints of primordial non-gaussianities on large-scale structure: scale dependent bias and abundance of virialized objects}",
    eprint = "0710.4560",
    archivePrefix = "arXiv",
    doi = "10.1103/PhysRevD.77.123514",
    journal = "Phys. Rev. D",
    volume = "77",
    pages = "123514",
    year = "2008"
}

@article{Matarrese_2008,
    author = "Matarrese, Sabino and Verde, Licia",
    title = "{The effect of primordial non-Gaussianity on halo bias}",
    eprint = "0801.4826",
    archivePrefix = "arXiv",
    doi = "10.1086/587840",
    journal = "Astrophys. J. Lett.",
    volume = "677",
    pages = "L77--L80",
    year = "2008"
}

@article{Liguori_2010,
    author = "Liguori, Michele and Sefusatti, Emiliano and Fergusson, James R. and Shellard, E. P. S.",
    title = "{Primordial non-Gaussianity and Bispectrum Measurements in the Cosmic Microwave Background and Large-Scale Structure}",
    eprint = "1001.4707",
    archivePrefix = "arXiv",
    doi = "10.1155/2010/980523",
    journal = "Adv. Astron.",
    volume = "2010",
    pages = "980523",
    year = "2010"
}

@misc{Alvarez_2014,
      title="{Testing Inflation with Large Scale Structure: Connecting Hopes with Reality}", 
      author={M. Alvarez et al.},
      year={2014},
      eprint={1412.4671},
      archivePrefix={arXiv},
      primaryClass={astro-ph.CO},
      url={https://arxiv.org/abs/1412.4671}, 
}

@article{dePutter_2014,
    author = "de Putter, Roland and Dor{\'e}, Olivier",
    title = "{Designing an Inflation Galaxy Survey: how to measure $\sigma(f_{\rm NL}) \sim 1$ using scale-dependent galaxy bias}",
    eprint = "1412.3854",
    archivePrefix = "arXiv",
    doi = "10.1103/PhysRevD.95.123513",
    journal = "Phys. Rev. D",
    volume = "95",
    number = "12",
    pages = "123513",
    year = "2017"
}

@article{DESI_2016,
    author = "Aghamousa, Amir and others",
    collaboration = "DESI",
    title = "{The DESI Experiment Part I: Science, Targeting, and Survey Design}",
    eprint = "1611.00036",
    archivePrefix = "arXiv",
    primaryClass = "astro-ph.IM",
    year = "2016"
}

@article{Dore_2014,
    author = "Dor{\'e}, Olivier and others",
    collaboration = "SPHEREx",
    title = "{Cosmology with the SPHEREx All-Sky Spectral Survey}",
    eprint = "1412.4872",
    archivePrefix = "arXiv",
    primaryClass = "astro-ph.CO",
    year = "2014"
}

@article{LSSTScienceBook_2009,
    author = "Abell, Paul A. and others",
    collaboration = "LSST Science, LSST Project",
    title = "{LSST Science Book, Version 2.0}",
    eprint = "0912.0201",
    archivePrefix = "arXiv",
    primaryClass = "astro-ph.IM",
    doi = "10.2172/1156415",
    year = "2009"
}

@article{Bernardeau_2002,
    author = "Bernardeau, F. and Colombi, S. and Gaztanaga, E. and Scoccimarro, R.",
    title = "{Large scale structure of the universe and cosmological perturbation theory}",
    eprint = "astro-ph/0112551",
    archivePrefix = "arXiv",
    doi = "10.1016/S0370-1573(02)00135-7",
    journal = "Phys. Rept.",
    volume = "367",
    pages = "1--248",
    year = "2002"
}

@article{Hamimeche_2008,
    author = "Hamimeche, Samira and Lewis, Antony",
    title = "{Likelihood Analysis of CMB Temperature and Polarization Power Spectra}",
    eprint = "0801.0554",
    archivePrefix = "arXiv",
    doi = "10.1103/PhysRevD.77.103013",
    journal = "Phys. Rev. D",
    volume = "77",
    pages = "103013",
    year = "2008"
}

@article{Upham_2020,
    author = "Upham, Robin E. and Whittaker, Lee and Brown, Michael L.",
    title = "{Exact joint likelihood of pseudo-$C_\ell$ estimates from correlated Gaussian cosmological fields}",
    eprint = "1908.00795",
    archivePrefix = "arXiv",
    doi = "10.1093/mnras/stz3225",
    journal = "Mon. Not. Roy. Astron. Soc.",
    volume = "491",
    number = "3",
    pages = "3165--3181",
    year = "2020"
}

@article{Bond_1991,
    author = "Bond, J. R. and Cole, S. and Efstathiou, G. and Kaiser, N.",
    title = "{Excursion set mass functions for hierarchical Gaussian fluctuations}",
    journal = "Astrophys. J.",
    volume = "379",
    pages = "440--460",
    year = "1991",
    doi = "10.1086/170520"
}

@article{LaceyCole_1993,
    author = "Lacey, Cedric and Cole, Shaun",
    title = "{Merger rates in hierarchical models of galaxy formation}",
    journal = "Mon. Not. Roy. Astron. Soc.",
    volume = "262",
    pages = "627--649",
    year = "1993",
    doi = "10.1093/mnras/262.3.627"
}

@book{DaleyVereJones_2003,
    author = "Daley, D. J. and Vere-Jones, D.",
    title = "{An Introduction to the Theory of Point Processes. Volume I: Elementary Theory and Methods}",
    edition = "2",
    publisher = "Springer",
    address = "New York",
    year = "2003"
}

@article{Desjacques_2009,
    author = "Desjacques, Vincent and Seljak, Uros and Iliev, Ilian",
    title = "{Scale-dependent bias induced by local non-Gaussianity: A comparison to N-body simulations}",
    eprint = "0811.2748",
    archivePrefix = "arXiv",
    primaryClass = "astro-ph",
    doi = "10.1111/j.1365-2966.2009.14721.x",
    journal = "Mon. Not. Roy. Astron. Soc.",
    volume = "396",
    pages = "85--96",
    year = "2009"
}

@article{Smith_2007,
    author = "Smith, Robert E. and Scoccimarro, Roman and Sheth, Ravi K.",
    title = "{The Scale Dependence of Halo and Galaxy Bias: Effects in Real Space}",
    eprint = "astro-ph/0609547",
    archivePrefix = "arXiv",
    doi = "10.1103/PhysRevD.75.063512",
    journal = "Phys. Rev. D",
    volume = "75",
    pages = "063512",
    year = "2007"
}

@article{Barreira_2020,
   title="{Galaxy bias and primordial non-Gaussianity: insights from galaxy formation simulations with IllustrisTNG}",
   volume={2020},
   ISSN={1475-7516},
   url={http://dx.doi.org/10.1088/1475-7516/2020/12/013},
   DOI={10.1088/1475-7516/2020/12/013},
   number={12},
   journal={Journal of Cosmology and Astroparticle Physics},
   publisher={IOP Publishing},
   author={Barreira, Alexandre and Cabass, Giovanni and Schmidt, Fabian and Pillepich, Annalisa and Nelson, Dylan},
   year={2020},
   month=Dec, pages={013–013} }

@article{Barreira_2022,
   title="{Predictions for local PNG bias in the galaxy power spectrum and bispectrum and the consequences for fnl constraints}",
   volume={2022},
   ISSN={1475-7516},
   url={http://dx.doi.org/10.1088/1475-7516/2022/01/033},
   DOI={10.1088/1475-7516/2022/01/033},
   number={01},
   journal={Journal of Cosmology and Astroparticle Physics},
   publisher={IOP Publishing},
   author={Barreira, Alexandre},
   year={2022},
   month=Jan, pages={033} }

@misc{perez2026impactgalaxyformationgalaxy,
      title="{The Impact of Galaxy Formation on Galaxy Biasing, and Implications for Primordial non-Gaussianity Constraints}", 
      author={Lucia A. Perez and Shy Genel and Elisabeth Krause and Rachel S. Somerville},
      year={2026},
      eprint={2602.04987},
      archivePrefix={arXiv},
      primaryClass={astro-ph.CO},
      url={https://arxiv.org/abs/2602.04987}, 
}

@article{Desjacques_2010,
   title="{Primordial Non‐Gaussianity in the Large‐Scale Structure of  the Universe}",
   volume={2010},
   ISSN={1687-7977},
   url={http://dx.doi.org/10.1155/2010/908640},
   DOI={10.1155/2010/908640},
   number={1},
   journal={Advances in Astronomy},
   publisher={Wiley},
   author={Desjacques, Vincent and Seljak, Uroš},
   editor={Huterer, Dragan},
   year={2010},
   month=Jan }

@misc{nguyen2026galaxysizescomplementaryzerobias,
      title="{Galaxy sizes as complementary (zero-)bias tracers of local primordial non-Gaussianity}", 
      author={Nhat-Minh Nguyen and Kazuyuki Akitsu and Atsushi Taruya},
      year={2026},
      eprint={2603.20196},
      archivePrefix={arXiv},
      primaryClass={astro-ph.CO},
      url={https://arxiv.org/abs/2603.20196}, 
}

@misc{barreira2023optimalrobustfrmnl,
      title="{Towards optimal and robust $f_{\rm NL}$ constraints with multi-tracer analyses}", 
      author={Alexandre Barreira and Elisabeth Krause},
      year={2023},
      eprint={2302.09066},
      archivePrefix={arXiv},
      primaryClass={astro-ph.CO},
      url={https://arxiv.org/abs/2302.09066}, 
}

@article{Pinon_2025,
   title="{A theoretical approach to density-split clustering}",
   volume={2025},
   ISSN={1475-7516},
   url={http://dx.doi.org/10.1088/1475-7516/2025/05/090},
   DOI={10.1088/1475-7516/2025/05/090},
   number={05},
   journal={Journal of Cosmology and Astroparticle Physics},
   publisher={IOP Publishing},
   author={Pinon, Mathilde and de Mattia, Arnaud and Burtin, Etienne and Ruhlmann-Kleider, Vanina and Codis, Sandrine and Paillas, Enrique and Cuesta-Lazaro, Carolina},
   year={2025},
   month=May, pages={090} }

@ARTICLE{1984ApJ,
       author = {{Kaiser}, N.},
        title = "{On the spatial correlations of Abell clusters.}",
      journal = {Astrophysical Journal},
         year = 1984,
        month = sep,
       volume = {284},
        pages = {L9-L12},
          doi = {10.1086/184341},
       adsurl = {https://ui.adsabs.harvard.edu/abs/1984ApJ...284L...9K}
}

@article{Bayer_2025,
   title="{The HalfDome multi-survey cosmological simulations: N-body simulations}",
   volume={2025},
   ISSN={1475-7516},
   url={http://dx.doi.org/10.1088/1475-7516/2025/05/016},
   DOI={10.1088/1475-7516/2025/05/016},
   number={05},
   journal={Journal of Cosmology and Astroparticle Physics},
   publisher={IOP Publishing},
   author={Bayer, Adrian E. and Zhong, Yici and Li, Zack and DeRose, Joseph and Feng, Yu and Liu, Jia},
   year={2025},
   month=May, pages={016} }

@article{Feng_2016,
   title="{FastPM: a new scheme for fast simulations of dark matter and haloes}",
   volume={463},
   ISSN={1365-2966},
   url={http://dx.doi.org/10.1093/mnras/stw2123},
   DOI={10.1093/mnras/stw2123},
   number={3},
   journal={Monthly Notices of the Royal Astronomical Society},
   publisher={Oxford University Press (OUP)},
   author={Feng, Yu and Chu, Man-Yat and Seljak, Uroš and McDonald, Patrick},
   year={2016},
   month=Aug, pages={2273–2286} }

@misc{chaussidon2025constrainingprimordialnongaussianitydesi,
      title="{Constraining primordial non-Gaussianity with DESI 2024 LRG and QSO samples}", 
      author={E. Chaussidon and C. Yèche and A. de Mattia and C. Payerne and P. McDonald and A. J. Ross and S. Ahlen and D. Bianchi and D. Brooks and E. Burtin and T. Claybaugh and A. de la Macorra and P. Doel and S. Ferraro and A. Font-Ribera and J. E. Forero-Romero and E. Gaztañaga and H. Gil-Marín and S. Gontcho A Gontcho and G. Gutierrez and J. Guy and K. Honscheid and C. Howlett and D. Huterer and R. Kehoe and D. Kirkby and T. Kisner and A. Kremin and L. Le Guillou and M. E. Levi and M. Manera and A. Meisner and R. Miquel and J. Moustakas and J. A. Newman and G. Niz and N. Palanque-Delabrouille and W. J. Percival and F. Prada and I. Pérez-Ràfols and C. Ravoux and G. Rossi and E. Sanchez and D. Schlegel and M. Schubnell and H. Seo and D. Sprayberry and G. Tarlé and M. Vargas-Magaña and B. A. Weaver and C. Zhao and H. Zou},
      year={2025},
      eprint={2411.17623},
      archivePrefix={arXiv},
      primaryClass={astro-ph.CO},
      url={https://arxiv.org/abs/2411.17623}, 
}

@article{Hand_2018,
   title="{nbodykit: An Open-source, Massively Parallel Toolkit for Large-scale Structure}",
   volume={156},
   ISSN={1538-3881},
   url={http://dx.doi.org/10.3847/1538-3881/aadae0},
   DOI={10.3847/1538-3881/aadae0},
   number={4},
   journal={The Astronomical Journal},
   publisher={American Astronomical Society},
   author={Hand, Nick and Feng, Yu and Beutler, Florian and Li, Yin and Modi, Chirag and Seljak, Uroš and Slepian, Zachary},
   year={2018},
   month=sep, pages={160} }

@article{Villaescusa_Navarro_2020,
   title="{The Quijote Simulations}",
   volume={250},
   ISSN={1538-4365},
   url={http://dx.doi.org/10.3847/1538-4365/ab9d82},
   DOI={10.3847/1538-4365/ab9d82},
   number={1},
   journal={The Astrophysical Journal Supplement Series},
   publisher={American Astronomical Society},
   author={Villaescusa-Navarro, Francisco and Hahn, ChangHoon and Massara, Elena and Banerjee, Arka and Delgado, Ana Maria and Ramanah, Doogesh Kodi and Charnock, Tom and Giusarma, Elena and Li, Yin and Allys, Erwan and Brochard, Antoine and Uhlemann, Cora and Chiang, Chi-Ting and He, Siyu and Pisani, Alice and Obuljen, Andrej and Feng, Yu and Castorina, Emanuele and Contardo, Gabriella and Kreisch, Christina D. and Nicola, Andrina and Alsing, Justin and Scoccimarro, Roman and Verde, Licia and Viel, Matteo and Ho, Shirley and Mallat, Stephane and Wandelt, Benjamin and Spergel, David N.},
   year={2020},
   month=Aug, pages={2} }

@MISC{Pylians,
    author = {{Villaescusa-Navarro}, Francisco},
    title = "{Pylians: Python libraries for the analysis of numerical simulations}",
    howpublished = {Astrophysics Source Code Library, record ascl:1811.008},
    year = 2018,
    month = nov,
    eid = {ascl:1811.008},
    pages = {ascl:1811.008},
    archivePrefix = {ascl},
    eprint = {1811.008},
    adsurl = {https://ui.adsabs.harvard.edu/abs/2018ascl.soft11008V}
}

@article{Tegmark_1997,
   title="{Karhunen‐Loeve Eigenvalue Problems in Cosmology: How Should We Tackle Large Data Sets?}",
   volume={480},
   ISSN={1538-4357},
   url={http://dx.doi.org/10.1086/303939},
   DOI={10.1086/303939},
   number={1},
   journal={The Astrophysical Journal},
   publisher={American Astronomical Society},
   author={Tegmark, Max and Taylor, Andy N. and Heavens, Alan F.},
   year={1997},
   month=May, pages={22–35} }

@article{Lewis_2000,
    author = "Lewis, Antony and Challinor, Anthony and Lasenby, Anthony",
    title = "{Efficient computation of CMB anisotropies in closed FRW models}",
    eprint = "astro-ph/9911177",
    archivePrefix = "arXiv",
    doi = "10.1086/309179",
    journal = "Astrophys. J.",
    volume = "538",
    pages = "473--476",
    year = "2000"
}

@misc{yuan2023desionepercentsurveyexploring,
      title="{The DESI One-Percent Survey: Exploring the Halo Occupation Distribution of Luminous Red Galaxies and Quasi-Stellar Objects with AbacusSummit}", 
      author={Sihan Yuan and Hanyu Zhang and Ashley J. Ross and Jamie Donald-McCann and Boryana Hadzhiyska and Risa H. Wechsler and Zheng Zheng and Shadab Alam and Violeta Gonzalez-Perez and Jessica Nicole Aguilar and Steven Ahlen and Davide Bianchi and David Brooks and Axel de la Macorra and Kevin Fanning and Jaime E. Forero-Romero and Klaus Honscheid and Mustapha Ishak and Robert Kehoe and James Lasker and Martin Landriau and Marc Manera and Paul Martini and Aaron Meisner and Ramon Miquel and John Moustakas and Seshadri Nadathur and Jeffrey A. Newman and Jundan Nie and Will Percival and Claire Poppett and Antoine Rocher and Graziano Rossi and Eusebio Sanchez and Lado Samushia and Michael Schubnell and Hee-Jong Seo and Gregory Tarle and Benjamin Alan Weaver and Jiaxi Yu and Zhimin Zhou and Hu Zou},
      year={2023},
      eprint={2306.06314},
      archivePrefix={arXiv},
      primaryClass={astro-ph.CO},
      url={https://arxiv.org/abs/2306.06314}, 
}

@article{Seljak_2009_2,
   title="{How to Suppress the Shot Noise in Galaxy Surveys}",
   volume={103},
   ISSN={1079-7114},
   url={http://dx.doi.org/10.1103/PhysRevLett.103.091303},
   DOI={10.1103/physrevlett.103.091303},
   number={9},
   journal={Physical Review Letters},
   publisher={American Physical Society (APS)},
   author={Seljak, Uroš and Hamaus, Nico and Desjacques, Vincent},
   year={2009},
   month=Aug }

@article{pymc2023,
  title = "{{PyMC}: A Modern and Comprehensive Probabilistic Programming Framework in {P}ython}",
  author = {Oriol Abril-Pla and Virgile Andreani and Colin Carroll and Larry Dong and Christopher J. Fonnesbeck and Maxim Kochurov and Ravin Kumar and Junpeng Lao and Christian C. Luhmann and Osvaldo A. Martin and Michael Osthege and Ricardo Vieira and Thomas Wiecki and Robert Zinkov },
  journal = {{PeerJ} Computer Science},
  volume = {9},
  number = {e1516},
  doi = {10.7717/peerj-cs.1516},
  year = {2023}
}

@misc{hoffman2011nouturnsampleradaptivelysetting,
      title="{The No-U-Turn Sampler: Adaptively Setting Path Lengths in Hamiltonian Monte Carlo}", 
      author={Matthew D. Hoffman and Andrew Gelman},
      year={2011},
      eprint={1111.4246},
      archivePrefix={arXiv},
      primaryClass={stat.CO},
      url={https://arxiv.org/abs/1111.4246}, 
}

@misc{sullivan2025localprimordialnongaussianbias,
      title="{Local Primordial Non-Gaussian Bias at the Field Level}", 
      author={James M. Sullivan and Shi-Fan Chen},
      year={2025},
      eprint={2410.18039},
      archivePrefix={arXiv},
      primaryClass={astro-ph.CO},
      url={https://arxiv.org/abs/2410.18039},
}

@article{Adame_2025,
   title="{DESI 2024 VII: cosmological constraints from the full-shape modeling of clustering measurements}",
   volume={2025},
   ISSN={1475-7516},
   url={http://dx.doi.org/10.1088/1475-7516/2025/07/028},
   DOI={10.1088/1475-7516/2025/07/028},
   number={07},
   journal={Journal of Cosmology and Astroparticle Physics},
   publisher={IOP Publishing},
   author={Adame, A.G. and Aguilar, J. and Ahlen, S. and Alam, S. and others},
   collaboration={DESI},
   year={2025},
   month=jul, pages={028} }

@misc{andrews2026fieldlevelinferenceprimordialnongaussianity,
      title="{Field-Level Inference of Primordial Non-Gaussianity with the Quijote Simulation Suite}", 
      author={Adam Andrews and Jens Jasche and Guilhem Lavaux and William Coulton and Francisco Villaescusa-Navarro and Marco Baldi and Drew Jamieson and Gabriel Jung and Dionysios Karagiannis and Florent Leclercq and Michele Liguori and Marco Marinucci and Benjamin Wandelt},
      year={2026},
      eprint={2603.20855},
      archivePrefix={arXiv},
      primaryClass={astro-ph.CO},
      url={https://arxiv.org/abs/2603.20855}, 
}

@misc{chen2026primordialnongaussianityfieldlevelcramerrao,
      title="{Primordial Non-Gaussianity and the Field-Level Cramer-Rao Bound}", 
      author={Eugene Chen and Daniel Green and Vincent S. H. Lee},
      year={2026},
      eprint={2603.22415},
      archivePrefix={arXiv},
      primaryClass={astro-ph.CO},
      url={https://arxiv.org/abs/2603.22415}, 
}

@misc{chudaykin2025reanalyzingdesidr13,
      title="{Reanalyzing DESI DR1: 3. Constraints on Inflation from Galaxy Power Spectra \& Bispectra}", 
      author={Anton Chudaykin and Mikhail M. Ivanov and Oliver H. E. Philcox},
      year={2025},
      eprint={2512.04266},
      archivePrefix={arXiv},
      primaryClass={astro-ph.CO},
      url={https://arxiv.org/abs/2512.04266}, 
}

@article{McDonald_2008,
   title={Primordial non-Gaussianity: Large-scale structure signature in the perturbative bias model},
   volume={78},
   ISSN={1550-2368},
   url={http://dx.doi.org/10.1103/PhysRevD.78.123519},
   DOI={10.1103/physrevd.78.123519},
   number={12},
   journal={Physical Review D},
   publisher={American Physical Society (APS)},
   author={McDonald, Patrick},
   year={2008},
   month=Dec }

@book{10.1373/clinchem.2003.025684,
  author    = {Anderson, T. W.},
  title     = {An Introduction to Multivariate Statistical Analysis},
  edition   = {3},
  year      = {2003},
  publisher = {John Wiley \& Sons},
  address   = {Hoboken, NJ},
  isbn      = {0-471-36091-0}
}

@article{Bock_2026,
   title="{The SPHEREx Satellite Mission}",
   volume={999},
   ISSN={1538-4357},
   url={http://dx.doi.org/10.3847/1538-4357/ae2be2},
   DOI={10.3847/1538-4357/ae2be2},
   number={1},
   journal={The Astrophysical Journal},
   publisher={American Astronomical Society},
   author="{Bock, James J. et al}",
   year={2026},
   month=Feb, pages={139} }

@article{Horowitz_2019,
   title={Efficient optimal reconstruction of linear fields and band-powers from cosmological data},
   volume={2019},
   ISSN={1475-7516},
   url={http://dx.doi.org/10.1088/1475-7516/2019/10/035},
   DOI={10.1088/1475-7516/2019/10/035},
   number={10},
   journal={Journal of Cosmology and Astroparticle Physics},
   publisher={IOP Publishing},
   author={Horowitz, B. and Seljak, U. and Aslanyan, G.},
   year={2019},
   month=Oct, pages={035–035} }

@article{Seljak_2017,
   title={Towards optimal extraction of cosmological information from nonlinear data},
   volume={2017},
   ISSN={1475-7516},
   url={http://dx.doi.org/10.1088/1475-7516/2017/12/009},
   DOI={10.1088/1475-7516/2017/12/009},
   number={12},
   journal={Journal of Cosmology and Astroparticle Physics},
   publisher={IOP Publishing},
   author={Seljak, Uroš and Aslanyan, Grigor and Feng, Yu and Modi, Chirag},
   year={2017},
   month=Dec, pages={009–009} }

@article{Seljak_1998,
doi = {10.1086/306019},
url = {https://doi.org/10.1086/306019},
year = {1998},
month = {aug},
publisher = {},
volume = {503},
number = {2},
pages = {492},
author = {Seljak, Uroš},
title = "{Cosmography and Power Spectrum Estimation: A Unified Approach}",
journal = {The Astrophysical Journal}
}

@article{Millea_2022,
   title="{Marginal unbiased score expansion and application to CMB lensing}",
   volume={105},
   ISSN={2470-0029},
   url={http://dx.doi.org/10.1103/PhysRevD.105.103531},
   DOI={10.1103/physrevd.105.103531},
   number={10},
   journal={Physical Review D},
   publisher={American Physical Society (APS)},
   author={Millea, Marius and Seljak, Uroš},
   year={2022},
   month=May }

@article{Coulton_2023,
   title={Quijote-PNG: Simulations of Primordial Non-Gaussianity and the Information Content of the Matter Field Power Spectrum and Bispectrum},
   volume={943},
   ISSN={1538-4357},
   url={http://dx.doi.org/10.3847/1538-4357/aca8a7},
   DOI={10.3847/1538-4357/aca8a7},
   number={1},
   journal={The Astrophysical Journal},
   publisher={American Astronomical Society},
   author={Coulton, William R and Villaescusa-Navarro, Francisco and Jamieson, Drew and Baldi, Marco and Jung, Gabriel and Karagiannis, Dionysios and Liguori, Michele and Verde, Licia and Wandelt, Benjamin D.},
   year={2023},
   month=Jan, pages={64} }

@article{10.1111/j.1365-2966.2009.15150.x,
    author = {Grossi, M. and Verde, L. and Carbone, C. and Dolag, K. and Branchini, E. and Iannuzzi, F. and Matarrese, S. and Moscardini, L.},
    title = {Large-scale non-Gaussian mass function and halo bias: tests on N-body simulations},
    journal = {Monthly Notices of the Royal Astronomical Society},
    volume = {398},
    number = {1},
    pages = {321-332},
    year = {2009},
    month = {09},
    issn = {0035-8711},
    doi = {10.1111/j.1365-2966.2009.15150.x},
    url = {https://doi.org/10.1111/j.1365-2966.2009.15150.x},
}

\newpage
\appendix
\section{Optimal Estimators for Bias Parameters}\label{app:OptEst}
Let $\delta_t$ be the overdensity field of some biased tracer. On large scales this field is well described as Gaussian~\cite{Bernardeau_2002}, so that its band-power estimates follow a Wishart distribution~\cite{Hamimeche_2008, Upham_2020}. In the following we derive optimal estimators for the bias parameters of $\delta_t$, assuming that the true dark matter field $\delta_m$ is known.

\subsection{Measuring $b_\phi$ from the Power Spectrum}
The data vector consists of the three spectra $(P_{tt}, P_{tm}, P_{mm})$ in various $k$ bins. Introducing the notation
\begin{equation}
    \mathbf X_i = N_i \begin{pmatrix}
        P_{tt}(k_i) & P_{tm}(k_i) \\
        P_{tm}(k_i) & P_{mm}(k_i)
    \end{pmatrix}, \qquad N_i = V \frac{k_i^2\,\Delta k}{4 \pi^2}
\end{equation}
for the \textit{observed} spectra and 
\begin{equation}
    \mathbf V_i = \begin{pmatrix}
        b_t(k_i)^2P_{mm}(k_i) + P_{\rm shot} & b_t(k_i)P_{mm}(k_i) \\
        b_t(k_i)P_{mm}(k_i) & P_{mm}(k_i)
    \end{pmatrix}, \qquad b_t(k_i) = b_1 + b_\phi \frac{\fnl}{\mathcal M(k_i)}
\end{equation}
for the \textit{predicted} spectra, the Wishart log-likelihood reads~\cite{Hamimeche_2008}
\begin{equation}
    \ell(\boldsymbol \theta ) = -\sum_i \left[\operatorname{tr}(\mathbf V_i^{-1}\mathbf X_i) + N_i\log \det \mathbf V_i\right], \qquad \boldsymbol \theta = (b_1, b_\phi, P_{\rm shot}).
\end{equation}
Assuming $\fnl = 0$ and solving $\partial \ell / \partial b_1 = 0$, one obtains the well-known estimator~\cite{Desjacques_2009, Smith_2007}
\begin{equation}
\label{eq:Estimator1}
    \hat b_1 = \frac{\sum_i \,N_iP_{tm}(k_i)}{\sum_i \, N_i P_{mm}(k_i)}.
\end{equation}
This is readily extended to the case where $\fnl \neq 0$ but is fixed to some known value. Then $\partial \ell / \partial b_1 = 0$ and $\partial \ell / \partial b_\phi = 0$ give the system of linear equations
\begin{equation}
    \begin{pmatrix}
    \sum_i N_i\, P_{mm}(k_i)/{\mathcal M(k_i)} & \sum_i N_i\,\fnl\,P_{mm}(k_i)/{\mathcal M(k_i)^2} \\ 
    \sum_i N_i\, P_{mm}(k_i)  &  \sum_i N_i\,\fnl\,P_{mm}(k_i)/{\mathcal M(k_i)}
    \end{pmatrix}
    \begin{pmatrix}
    \hat b_1 \\ \hat b_\phi
    \end{pmatrix} = \begin{pmatrix}
    \sum_i N_i P_{tm}(k_i)/\mathcal M(k_i) \\
    \sum_i N_i P_{tm}(k_i)
    \end{pmatrix}. \nonumber
\end{equation}
We introduce the notation $S_n \equiv \sum_i N_i\, P_{mm}(k_i)/{\mathcal M(k_i)^n}$ and $R_n \equiv \sum_i N_i\, P_{tm}(k_i)/{\mathcal M(k_i)^n}$ to obtain the estimators
\begin{equation}
    \label{eq:Estimator2}
    \hat b_1 = \frac{S_1R_1 - S_2R_0}{S_1^2 - S_0S_2}, \qquad \hat b_\phi = \frac{1}{\fnl}\frac{S_1R_0 - S_0R_1}{S_1^2 - S_0S_2}.    
\end{equation}

\subsection{Measuring $b_\phi$ from the Field-Level Difference}
We now consider two simulations run with different values of $\fnl$, namely $\fnl^{(+)}$ and $\fnl^{(-)}$, but sharing the same initial-condition seed. Letting $\delta_t^{(+)}$ and $\delta_t^{(-)}$ denote the corresponding tracer overdensity fields, and noting that the linear-bias term $b_1\delta_m$ is identical in the two runs, the \textit{difference field} takes the form
\begin{equation}
    \Delta(k) \equiv \delta_t^{(+)}(k)- \delta_t^{(-)}(k) = \delta\fnl \,b_\phi\,\phi(k) + \varepsilon + \mathcal O\left( (\delta\fnl)^2 \right),
\end{equation}    
where $\delta\fnl = \fnl^{(+)} - \fnl^{(-)}$. Correlating the above with the shared dark matter field $\delta_m$, we can make use of the optimal estimator~(\ref{eq:Estimator1}) to derive the optimal $b_\phi$ estimator
\begin{equation}
    \hat b_\phi = \frac{1}{\delta \fnl}\frac{\sum_i \, N_iP_{\Delta m}(k_i)/\mathcal M(k_i)}{\sum_i \,N_i P_{mm}(k_i)/\mathcal{M}(k_i)^2}.\label{eq:Estimator3}
\end{equation}

\section{Conditional Dark-Matter Environment of Halos}\label{app:CondDist}
In this appendix we derive the conditional probability $p(\delta)$ of finding the smoothed matter environment $(\delta_m)_R$ in an infinitesimal bin around $\delta$ given halo formation, which is used in Section~\ref{sec:TheoBack}. We work within the excursion-set framework and consider, for concreteness, halos in a narrow mass bin around $M$. In excursion-set theory one models $(\delta_m)_R$ as a Brownian trajectory in the variance $S_R = \sigma^2(R) = \int_{\boldsymbol k}P(k)W_R^2(k)$, starting from zero on the largest scales~\cite{Bond_1991, LaceyCole_1993}, with halo formation identified with the first crossing of the collapse threshold $\delta_c$. Taking the environment scale large enough that $S_M \gg S_R$, the density smoothed on scale $R$ conditioned on the formation of a halo of mass $M$ can be written, via Bayes' theorem, as
\begin{equation}
    p((\delta_m)_R | \text{first crossing at }S_M) = \frac{p((\delta_m)_R, \text{ no crossing in }[0, S_R])\cdot p(\text{first crossing at }S_M\,|\,(\delta_m)_R)}{p(\text{first crossing at }S_M)}. \nonumber
\end{equation}
Using the properties of Brownian motion the first piece becomes
\begin{equation}
    p((\delta_m)_R, \text{ no crossing in }[0, S_R]) = \frac{1}{\sqrt{2 \pi S_R}}\left(e^{-[(\delta_m)_R]^2/2S_R} - e^{-(2\delta_c - (\delta_m)_R)^2/2S_R} \right),
\end{equation}
where $\delta_c$ is the collapse threshold; see e.g.\ \cite{Desjacques_2018}. The second factor is the conditional first-crossing distribution \cite{Desjacques_2018}
\begin{equation}
    p(\text{first crossing at }S_M\,|\,(\delta_m)_R) = \frac{1}{\sqrt{2 \pi}}\frac{\delta_c - (\delta_m)_R}{(S_M - S_R)^{3/2}}\exp\left( -\frac{(\delta_c - (\delta_m)_R)^2}{2(S_M - S_R)}\right).
\end{equation}
Finally, $p(\text{first crossing at }S_M)$ is the unconditional first-crossing distribution
\begin{equation}
    p(\text{first crossing at }S_M) = \frac{d}{dS_M}F(>M) = \frac{\delta_c}{\sqrt{2 \pi S_M^3}}\exp\left(-\frac{\delta_c^2}{2S_M} \right), 
\end{equation}
where $F(> M) = \mathrm{erfc}(\delta_c/\sqrt{2S_M})$ is the standard Press-Schechter collapse fraction, see e.g.~\cite{Desjacques_2018}. We now combine these factors and take the limits $S_M, \delta_c^2 \gg S_R$ to find
\begin{equation}
    p((\delta_m)_R\,|\,\text{a halo of mass }M\text{ formed}) \approx \frac{1}{\sqrt{2 \pi S_R}}\left(1 - \frac{(\delta_m)_R}{\delta_c} \right)\exp\left(-\frac{((\delta_m)_R - \delta_c\,S_R/S_M)^2}{2S_R} \right).
\end{equation}
Neglecting the $(\delta_m)_R/\delta_c$ correction in the prefactor, this reduces to a Gaussian
\begin{equation}
    p(\delta) \approx \frac{1}{\sigma\sqrt{2\pi}}\exp\!\left(-\frac{(\delta-\mu)^2}{2\sigma^2}\right)d\delta, \qquad \mu = \frac{\delta_c S_R}{S_M}, \quad \sigma = \sqrt{S_R},
\end{equation}
confirming the claim in Section~\ref{sec:TheoBack} that $\mu$ is independent of $\sigma_8$ and $\sigma \propto \sigma_8$.

As $(\delta_m)_R \to \delta_c$ the prefactor $1 - (\delta_m)_R/\delta_c$ contributes meaningfully and the Gaussian approximation breaks down. For sufficiently large smoothing radii, however, the relevant range of $(\delta_m)_R$ stays well below $\delta_c$ and this tail can be ignored. Moreover, since the prefactor is independent of $\sigma_8$, it does not contribute to $b_\phi$.

\section{Halo-Environment Split: Separate-universe Derivation of $b_\phi(\delta_h)$}\label{app:b_phi_of_dh}
We now split a halo sample by the value of its own smoothed overdensity field
$(\delta_h)_R$, evaluated at the halo positions with a Gaussian kernel of radius $R$. We model the smoothed halo field as
\begin{equation}
    (\delta_h)_R = b_1\,(\delta_m)_R + (\varepsilon)_R, \qquad \langle  (\varepsilon)_R  (\varepsilon)_R \rangle = \sigma_{\varepsilon}^2 = \frac{1}{\bar n}\int_{\boldsymbol k}|W_R|^2 = \frac{1}{\bar n}\frac{1}{8 \pi ^{3/2}R^3}
\end{equation}
where the last equality assumes a Gaussian smoothing kernel. Conditioned on the halo positions we take $(\delta_m)_R \sim \mathcal N(\mu, \sigma^2)$ for the matter contribution, and claim that the stochastic piece obeys $(\varepsilon)_R \sim \mathcal N(\mu_\varepsilon, \sigma_{\varepsilon}^2)$ with a non-zero mean $\mu_\varepsilon = W_R(0)/\bar n = \big((2\pi)^{3/2}R^3\bar n\big)^{-1}$. To see the latter, write the number-density field as $n(\boldsymbol x) = \sum_a \delta_D(\boldsymbol x - \boldsymbol x_{(a)})$; convolving with $W_R$ gives $(\delta_h)_R(\boldsymbol x) = \bar n^{-1}\sum_a W_R(\boldsymbol x - \boldsymbol x_{(a)}) - 1$, so that evaluated at a halo position $\boldsymbol x_{(i)}$,
\begin{equation}
    (\delta_h)_R(\boldsymbol x_{(i)}) = \frac{W_R(0)}{\bar n} + \frac{1}{\bar n}\sum_{a\neq i}W_R(\boldsymbol x_{(i)} - \boldsymbol x_{(a)}) - 1.
\end{equation}
By Slivnyak's theorem, conditioning on a halo at $\boldsymbol x_{(i)}$ leaves the statistics of the remaining points $\{a \neq i\}$ unchanged~\cite{DaleyVereJones_2003, daley2008introduction}, so the sum $\bar n^{-1}\sum_{a\neq i}W_R(\boldsymbol x_{(i)} - \boldsymbol x_{(a)}) - 1$ has the same zero-mean noise contribution as $(\varepsilon)_R$ at a random point. The term $W_R(0)/\bar n$ is a deterministic constant, which shifts the conditional mean of $(\varepsilon)_R$ to $\mu_\varepsilon$ without affecting its variance.
Adding the matter and stochastic contributions, the smoothed halo overdensity at the halo positions is Gaussian,
\begin{equation}
    (\delta_h)_R | \text{halo positions} \sim \mathcal N(b_1 \mu + \mu_\varepsilon, b_1 ^2\sigma^2 +  \sigma_{\varepsilon}^2).
\end{equation}
The $\fnl$ response of the selected sub-sample follows from differentiating its abundance with respect to $\sigma_8$. The probability $p(\delta_h) \equiv p(\delta_h\,|\,\text{halo positions})$, where we set $\delta_h \equiv (\delta_h)_R $ to compactify the notation, is Gaussian with mean $\mu_h$ and width $\sigma_h$,
\begin{equation}
    \log p(\delta_h) = \mathrm{const} - \log \sigma_h - \frac{(\delta_h - \mu_h)^2}{2\sigma_h^2},
\end{equation}
so that, defining $\nu \equiv (\delta_h - \mu_h)/\sigma_h$, its derivatives with respect to the conditional mean and width are
\begin{equation}
    \frac{\partial \log p(\delta_h)}{\partial \mu_h} = \frac{\nu}{\sigma_h}, \qquad \frac{\partial \log p(\delta_h)}{\partial \log \sigma_h} = \nu^2 - 1 .
\end{equation}
The excess PNG bias of the sub-sample is $\Delta b_\phi = 2\,\partial \log p(\delta_h)/\partial \log \sigma_8$. Applying the chain rule through $\mu_h$ and $\sigma_h$ gives
\begin{equation}
    \Delta b_\phi = 2\frac{\partial \mu_h}{\partial \log \sigma_8}\frac{\partial \log p(\delta_h)}{\partial \mu_h} + 2\frac{\partial \log \sigma_h}{\partial \log \sigma_8}\frac{\partial \log p(\delta_h)}{\partial \log \sigma_h} = \frac{2}{\sigma_h}\frac{\partial \mu_h}{\partial \log \sigma_8}\,\nu + \frac{\partial \log \sigma_h^2}{\partial \log \sigma_8}\,(\nu^2 - 1),
\end{equation}
where $\mu_h = b_1 \mu + \mu_\varepsilon$ and $\sigma_h^2 = b_1 ^2\sigma^2 + \sigma_{\varepsilon}^2$ are the conditional mean and variance found above. It remains to evaluate the two $\sigma_8$-derivatives. The matter mean $\mu = \delta_c S_R/S_M$ is a ratio of variances and hence independent of $\sigma_8$. Introducing the notation $\partial b_1/\partial \log \sigma_8 = b_1\gamma$ and noting $\partial \mu_\varepsilon/\partial \log \sigma_8 = -\mu_\varepsilon\,\partial \log \bar n/\partial \log \sigma_8 = -\tfrac{1}{2}\mu_\varepsilon b_\phi$, we find
\begin{equation}
    \frac{\partial \mu_h}{\partial \log \sigma_8} = b_1\gamma\,\mu - \tfrac{1}{2}\mu_\varepsilon b_\phi.
\end{equation}
Likewise, using $\sigma \propto \sigma_8$ (so $\partial \sigma^2/\partial \log \sigma_8 = 2\sigma^2$) and $\sigma_\varepsilon^2 \propto 1/\bar n$,
\begin{equation}
    \frac{\partial \sigma_h^2}{\partial \log \sigma_8} = \frac{\partial (b_1^2\sigma^2)}{\partial \log \sigma_8} + \frac{\partial \sigma_\varepsilon^2}{\partial \log \sigma_8} = 2 b_1^2\sigma^2(1+\gamma) - \frac{1}{2}\sigma_\varepsilon^2 b_\phi .
\end{equation}
The two derivatives combine into the excess PNG bias of the sub-sample selected at $\delta_h$,
\begin{equation}
    \Delta b_\phi(\delta_h) = A\,\nu - B \left(1 - \nu^2\right), \qquad \nu = \frac{\delta_h - \mu_h}{\sigma_h},
\end{equation}
with the slope and curvature coefficients
\begin{equation}
A \equiv \frac{2}{\sigma_h}\frac{\partial \mu_h}{\partial \log \sigma_8} = \frac{2\,b_1\,\mu\,\gamma - \mu_\varepsilon\, b_\phi}{\sigma_h},
\quad
B \equiv \frac{\partial \log \sigma_h^2}{\partial \log \sigma_8} = 2\left(\,\frac{b_1 ^2\sigma ^2}{\sigma _h ^2}(1+\gamma) - \frac{1}{4}\frac{\sigma_\varepsilon^2}{\sigma_h^2}\,b_\phi\,\right).
\end{equation}

\section{Gaussian Conditioning}\label{app:GaussCond}
Let $\zeta(\boldsymbol x)$ be a Gaussian random field with
\begin{equation}
    \langle \zeta(\boldsymbol x) \rangle = 0, \qquad
    \langle \zeta(\boldsymbol x)\,\zeta(\boldsymbol y)\rangle = \xi(\boldsymbol x,\boldsymbol y), \qquad
    \sigma^2 \equiv \xi(\boldsymbol x,\boldsymbol x).
\end{equation}
We take the number-density field to be
\begin{equation}
    n(\boldsymbol x) = n_0\{ 1+\zeta(\boldsymbol x)\}.
\end{equation}
We then construct a selected subsample $\eta(\boldsymbol x)$ by retaining only those locations where the Gaussian field takes the value
$\zeta(\boldsymbol x)=\zeta_0$.

A convenient way to represent this selection is through a delta-function constraint. Weighting the underlying field $n(\boldsymbol x)$ with a Dirac delta that enforces $\zeta(\boldsymbol x)=\zeta_0$ gives
\begin{equation}
    n_\eta(\boldsymbol x)\, d\zeta_0
    \;\propto\;
    \big(1+\zeta(\boldsymbol x)\big)\,
    \delta_D\!\big(\zeta(\boldsymbol x)-\zeta_0\big)\, d\zeta_0
    \;=\;
    (1+\zeta_0)\,
    \delta_D\!\big(\zeta(\boldsymbol x)-\zeta_0\big)\, d\zeta_0 .
\end{equation}
Taking the average, and using $\langle \delta_D\!\big(\zeta(\boldsymbol x)-\zeta_0\big) \rangle = p(\zeta_0)$, we obtain
\begin{equation}
    \langle n_\eta(\boldsymbol x)\rangle\, d\zeta_0
    \;\propto\;
    (1+\zeta_0)\,p(\zeta_0)\, d\zeta_0 .
\end{equation}
We now define the subsample overdensity $\eta(\boldsymbol x)$ in the standard way through
\begin{equation}
    n_\eta(\boldsymbol x) \equiv \langle n_\eta\rangle\,\big(1+\eta(\boldsymbol x)\big).
\end{equation}
With the above proportionalities, the factor $(1+\zeta_0)$ cancels, leading to
\begin{equation}
    1+\eta(\boldsymbol x)
    =
    \frac{\delta_D\!\big(\zeta(\boldsymbol x)-\zeta_0\big)}{p(\zeta_0)}
    \qquad\Rightarrow\qquad
    \eta(\boldsymbol x)=\frac{\delta_D\!\big(\zeta(\boldsymbol x)-\zeta_0\big)}{p(\zeta_0)}-1.
    \label{eq:eta_def}
\end{equation}
We are interested in the cross-correlation between the selected overdensity $\eta(\boldsymbol x)$ and the underlying Gaussian field $\zeta(\boldsymbol y)$. Since $\langle \zeta(\boldsymbol y)\rangle=0$, the constant ``$-1$'' in Eq.~\eqref{eq:eta_def} does not contribute, and we have
\begin{align}
    \langle \eta(\boldsymbol x)\,\zeta(\boldsymbol y)\rangle
    &=
    \frac{1}{p(\zeta_0)}
    \Big\langle
    \delta_D\!\big(\zeta(\boldsymbol x)-\zeta_0\big)\,\zeta(\boldsymbol y)
    \Big\rangle .
    \label{eq:cross_start}
\end{align}
To evaluate the remaining expectation value, introduce $\zeta_x\equiv \zeta(\boldsymbol x)$ and $\zeta_y\equiv \zeta(\boldsymbol y)$. Writing the expectation in terms of the joint Gaussian PDF $p(\zeta_x,\zeta_y)$ gives
\begin{align}
    \Big\langle
    \delta_D\!\big(\zeta(\boldsymbol x)-\zeta_0\big)\,\zeta(\boldsymbol y)
    \Big\rangle
    &=
    \int d\zeta_x\, d\zeta_y\;
    p(\zeta_x,\zeta_y)\,
    \delta_D(\zeta_x-\zeta_0)\,\zeta_y
    \nonumber\\
    &=
    \int d\zeta_y\;
    p(\zeta_0,\zeta_y)\,\zeta_y .
    \label{eq:joint_int}
\end{align}
We then factorize the joint PDF as $p(\zeta_0,\zeta_y)=p(\zeta_0)\,p(\zeta_y|\zeta_0)$ to obtain
\begin{align}
    \int d\zeta_y\; p(\zeta_0,\zeta_y)\,\zeta_y
    &=
    p(\zeta_0)\,
    \mathbb{E}\!\left[\zeta(\boldsymbol y)\,\big|\,\zeta(\boldsymbol x)=\zeta_0\right].
\end{align}
Substituting into Eq.~\eqref{eq:cross_start}, the factors of $p(\zeta_0)$ cancel, and one arrives at
\begin{equation}
    \langle \eta(\boldsymbol x)\,\zeta(\boldsymbol y)\rangle
    =
    \mathbb{E}\!\left[\zeta(\boldsymbol y)\,\big|\,\zeta(\boldsymbol x)=\zeta_0\right].
    \label{eq:cross_condmean}
\end{equation}
For a bivariate Gaussian $(\zeta_x,\zeta_y)$ with covariance matrix
\begin{equation}
    \Sigma =
    \begin{pmatrix}
        \sigma^2 & \xi(\boldsymbol x,\boldsymbol y) \\
        \xi(\boldsymbol x,\boldsymbol y) & \sigma^2
    \end{pmatrix},
\end{equation}
the conditional mean is given by the standard Gaussian formula~\cite{10.1373/clinchem.2003.025684}
\begin{equation}
    \mathbb{E}\!\left[\zeta(\boldsymbol y)\,\big|\,\zeta(\boldsymbol x)=\zeta_0\right]
    =
    \frac{\xi(\boldsymbol x,\boldsymbol y)}{\sigma^2}\,\zeta_0.
\end{equation}
Combining with Eq.~\eqref{eq:cross_condmean} yields the final result
\begin{equation}
    \langle \eta(\boldsymbol x)\,\zeta(\boldsymbol y)\rangle
    =
    \frac{\zeta_0}{\sigma^2}\,\langle \zeta(\boldsymbol x)\,\zeta(\boldsymbol y)\rangle.
    \label{eq:final_cross}
\end{equation}
This shows that conditioning (or splitting) a Gaussian field by the \emph{same} Gaussian field rescales the two-point function. In the context of biased tracers, this rescaling argument implies that such a conditioning induces an overall rescaling of the corresponding (full, scale-dependent) bias response of the selected sample. That is $b_1, b_\phi$ are rescaled by the same factor so that $b_1/b_\phi$ remains constant. 

\section{Inference for the Multitracer Fit}\label{app:Fitting}
Here we describe the procedure used to infer $f_{\mathrm{NL}}$ from a multitracer analysis where the halo sample is split by its large-scale dark matter environment. Unless stated otherwise, all results shown in this appendix are presented for the environmental split based on the reconstructed matter field $\hat\delta_m$, using Eq.~\eqref{eq:Tranf}. However, the same inference pipeline applies equally well if one instead uses the true $(\delta_m)_R$ (and our validation tests confirm that both approaches give unbiased $f_{\mathrm{NL}}$ constraints).

\begin{figure}
    \centering
    \includegraphics[width=1\linewidth]{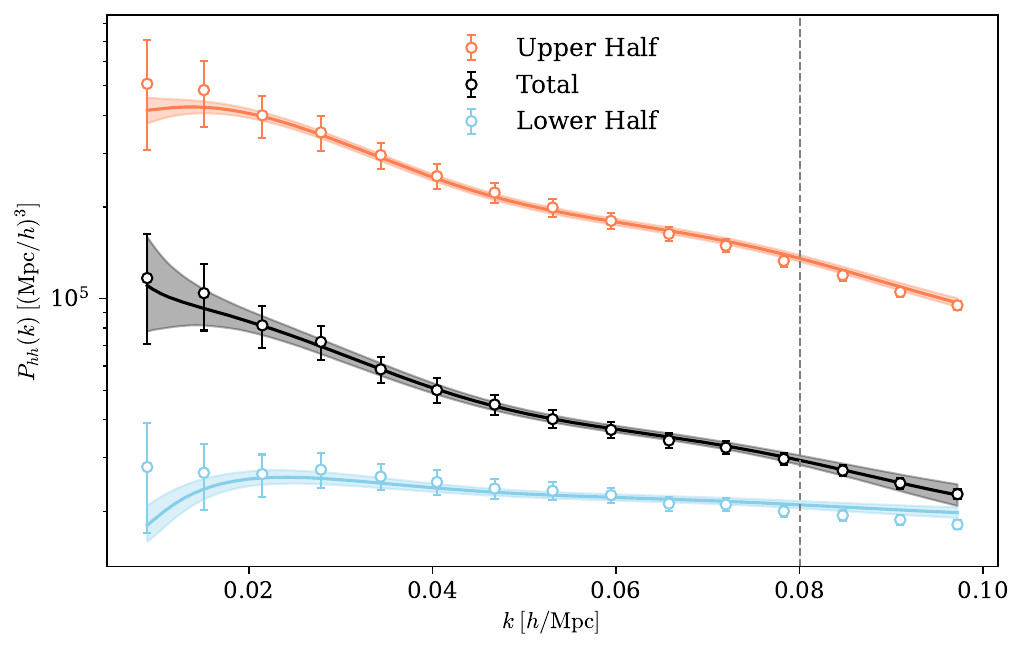}
    \caption{Power spectra of the tracers considered in this appendix.
    The total sample is shown in black, whereas the split tracers are shown in color:
    the low-density environments in skyblue and the high-density subsample in coral.
    Posterior-based fits are also indicated.}
    \label{fig:PggSplit}
\end{figure}

\begin{table}
    \centering
    \begin{tabular}{c|cccccc}
        variable & $b_1^{\rm high}$ & $b_1^{\rm low}$ & $f_{\mathrm{NL}}$ & $s^{\rm low}$ & $s^{\rm high}$ & $s_\times$ \\ \hline
        prior & $\mathcal U([0, 10])$ & $\mathcal U([-10, 0])$ & $\mathcal U([-500, 500])$
               & $\mathcal N(1, 1^2)_{>0}$
               & $\mathcal N(0.1, 0.5^2)_{>0}$
               & $\mathcal N(1, 1^2)$ \\
    \end{tabular}
    \caption{Priors chosen for the multitracer $f_{\mathrm{NL}}$ inference.
    Since $\Delta b_1 \sim 3$, we typically have $b_1^{\rm low}<0$ and enforce this in the prior.
    Moreover, more massive halos have lower shot noise \cite{Seljak_2009_2} and preferentially live in high-density regions,
    so we expect $s^{\rm high}$ to be close to zero.
    The cross stochasticity $s_\times$ is given a broad uninformative prior.}
    \label{tab:Prior}
\end{table}

We split the tracer population into two equally sized environment bins and adopt the key theoretical assumptions of the work:
($i$) we set $\Delta b_\phi = 0$ throughout the split (as motivated in Sec.~\ref{sec:TheoBack}), and
($ii$) we fix the PNG-response parameter to the universality prediction evaluated for the \emph{parent} (unsplit) halo sample,
$b_\phi = 2\delta_c\,(b_1-1)$, where we use $\delta_c = \sqrt{0.75} \cdot 1.686$ as appropriate for \texttt{FoF} halos \cite{Desjacques_2010, sullivan2025localprimordialnongaussianbias, 10.1111/j.1365-2966.2009.15150.x}.
The free parameters of the fit are the linear biases of the two subsamples, $b_1^{\rm high}$ and $b_1^{\rm low}$, the global $f_{\mathrm{NL}}$, nuisance parameters describing deviations from Poisson shot noise, $s^{\rm high}$ and $s^{\rm low}$, and a cross stochasticity $s_\times$ parameterizing a non-Poissonian contribution to the off-diagonal of the covariance.

The tracer is defined as halos at redshift $z=1$ in the 10 \texttt{Quijote-PNG} simulations, using a mass cut $M_{\min}=1.3\times 10^{13}\,M_\odot/h$.
While Sec.~\ref{sec:TheoBack} strictly considered halos in a narrow mass bin (for which $\Delta b_\phi=0$ holds exactly), the mass cut only realizes this condition approximately.
Because our validation tests recover $f_{\mathrm{NL}}$ without bias, we expect the impact of this distinction to be small.
In an observational analysis, this approximation can be mitigated by first splitting the sample into mass bins and then applying the environmental split.

We average over the 10 simulation seeds to reduce statistical noise.
As in Appendix~\ref{app:OptEst}, we model the distribution of large-scale power-spectrum estimators using a Wishart likelihood.
Accordingly, the log-likelihood reads
\begin{equation}
    \ell(\boldsymbol \theta ) =
    -\sum_i \left[\operatorname{tr}(\mathbf V_i^{-1}\mathbf X_i)
    + N_i\log \det \mathbf V_i\right],
    \qquad
    \boldsymbol \theta = (b_1^{\rm high}, b_1^{\rm low}, f_{\mathrm{NL}}, s^{\rm low}, s^{\rm high}, s_\times),
\end{equation}
where the sum runs over the $k$-bins used in the fit and we take $k_{\max} \equiv 0.08\,h/\mathrm{Mpc}$.

For the two-environment (low/high) case, we model the expected $2\times 2$ covariance matrix of the power-spectrum estimators as
\begin{equation}
    \mathbf V_i =
    \begin{pmatrix}
        b^{\rm low}(k_i)^2P_{mm}(k_i) + 2\,{s^{\rm low}}/{\bar n}
        &
        b^{\rm low}(k_i)b^{\rm high}(k_i) P_{mm}(k_i) + {s_\times}/{\bar n}
        \\
        b^{\rm low}(k_i)b^{\rm high}(k_i)P_{mm}(k_i) + {s_\times}/{\bar n}
        &
        b^{\rm high}(k_i)^2P_{mm}(k_i) + 2\,{s^{\rm high}}/{\bar n}
    \end{pmatrix},
\end{equation}
with
\begin{equation}
    b^{(i)}(k) = b_1^{(i)} + b_\phi \frac{f_{\mathrm{NL}}}{\mathcal M(k)} ,
\end{equation}
where $b_\phi$ and $\bar n$ are taken from the parent (unsplit) population.
Priors for all parameters are specified in Tab.~\ref{tab:Prior}.

\begin{figure}
    \centering
    \includegraphics[width=1\linewidth]{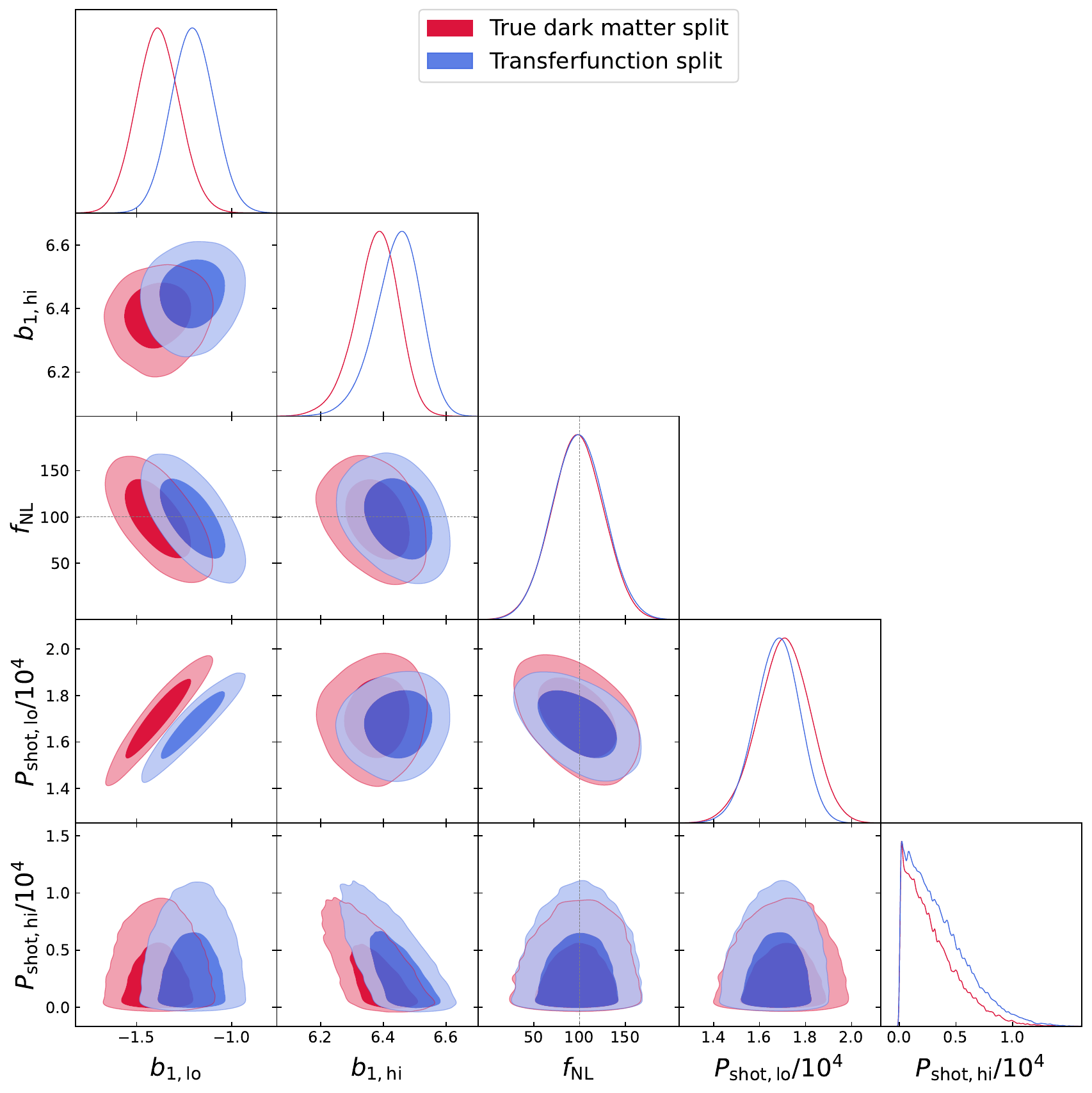}
    \caption{Corner plot of the posterior samples for the multitracer $f_{\mathrm{NL}}$ fit, comparing the split based on the \emph{true} dark matter environment $(\delta_m)_R$ (red) with the split based on its linear-theory estimate $\hat\delta_m$ from Eq.~\eqref{eq:Tranf} (blue).
    Both recover the input value $f_{\mathrm{NL}}=100$ (dashed line) without bias.
    In both cases the cross stochasticity $s_\times$ is included as a free parameter and marginalized over.
    We introduced the notation $P_{\rm shot,(i)} = 2\,s^{(i)}/\bar n$.}
    \label{fig:Corner}
\end{figure}

Contrary to the situation studied in Appendix~\ref{app:OptEst}, this posterior is no longer analytically tractable because the likelihood depends on $f_{\mathrm{NL}}$ and the nuisance parameters $s^{\rm low/high}$ in a coupled way.
We therefore implemented the Wishart likelihood in \texttt{PyMC} \cite{pymc2023} and sampled with NUTS \cite{hoffman2011nouturnsampleradaptivelysetting}.
A corner plot of the posterior samples is presented in Fig.~\ref{fig:Corner}, where we overlay the fit based on the true dark matter environment with the fit based on its transfer-function estimate.
In the plot we use the notation $P_{\rm shot,(i)} \equiv 2\,s^{(i)}/\bar n$ to make the physical meaning of the shot-noise parameters explicit.

The environmental split is a non-trivial selection, so the resulting sub-samples need not inherit the simple Poisson stochasticity of the parent population.
We therefore include the cross stochasticity $s_\times$ as a free parameter and marginalize over it for both splits.
We find $s_\times \neq 0$ at high significance; neglecting it biases $f_{\mathrm{NL}}$ low.
Higher-order, scale-dependent ($\propto k^2$) stochastic counterterms, which are standard in full-shape clustering analyses \cite{Adame_2025}, are not included in the analysis, as we find no evidence for them at our redshift and $k_{\max}$.

\newpage
\section{Further Forecasts}\label{app:FurtherFisher}

In the main text, we considered the most immediate application of the environmental-split method to current data: dividing the relatively high-density LRG sample into two equal-sized subpopulations. The simplicity of this binary split makes it particularly promising for observational applications. Figure~\ref{fig:Fisher} shows that substantial multitracer gains emerge once the number density reaches approximately
$\bar n \sim 10^{-4}\,(h/\mathrm{Mpc})^3$.
In this appendix, we investigate whether additional improvements can be obtained by combining environmental information with proxies for halo assembly history. We also consider an application to the DESI quasar (QSO) sample, which currently provides the tightest LSS constraints on $\fnl$.

\begin{figure}[htbp]
    \centering
    \begin{subfigure}[b]{1.0\textwidth}
        \centering
        \includegraphics[width=0.7\linewidth]{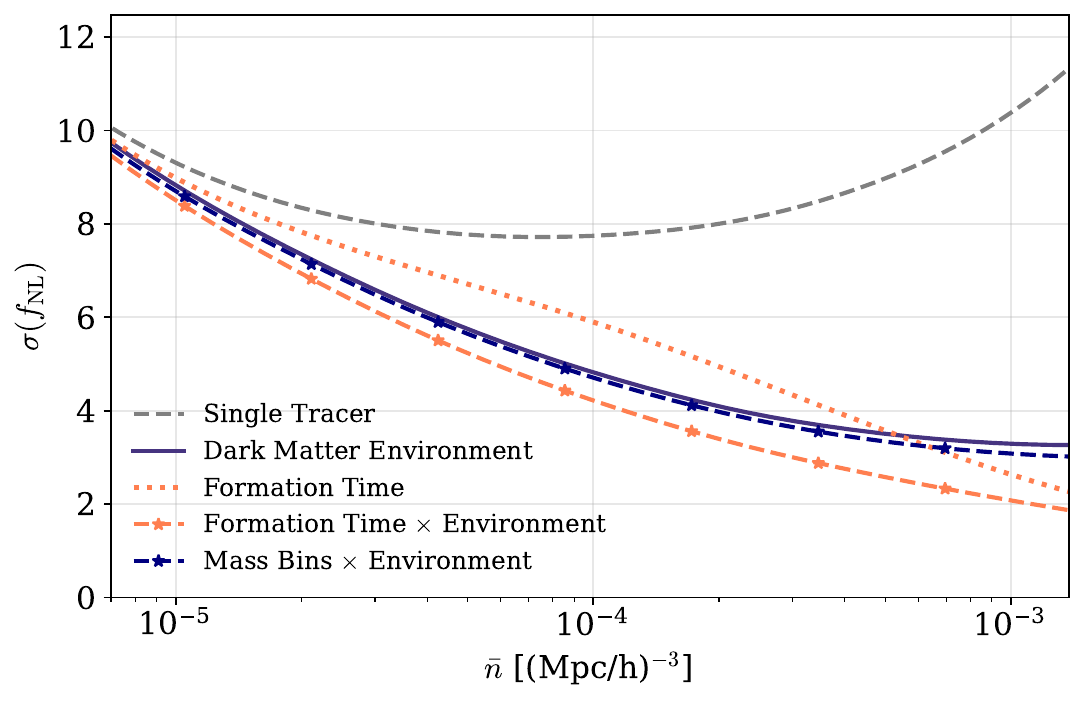}
        \includegraphics[width=0.7\linewidth]{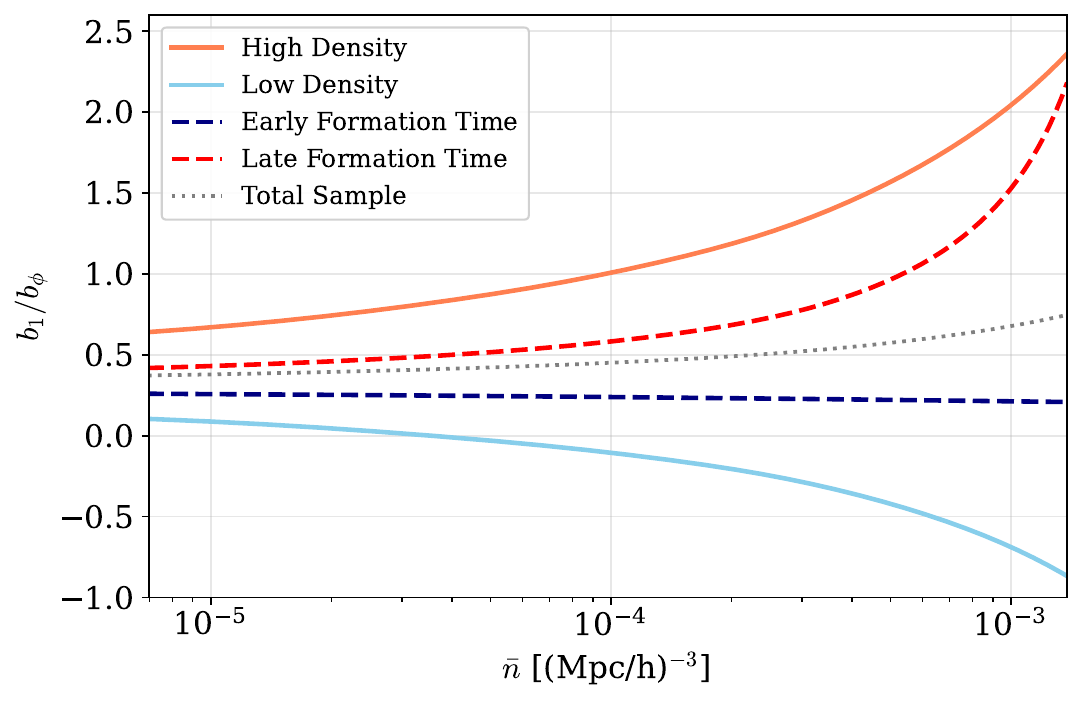}
        \caption{Forecasted improvements obtained by combining density- or halo-mass-based binning with the density split (\textit{upper panel}) and an illustration of the extremization of the Fisher information~\eqref{eq:ScalingLaw} achieved by these splits (\textit{lower panel}). See text for further details.}
        \label{fig:FurtherForecasts}
    \end{subfigure}
    \\ 
    \vspace{0.5cm}
    \begin{subfigure}[b]{1.0\textwidth}
        \centering
        \includegraphics[width=0.75\linewidth]{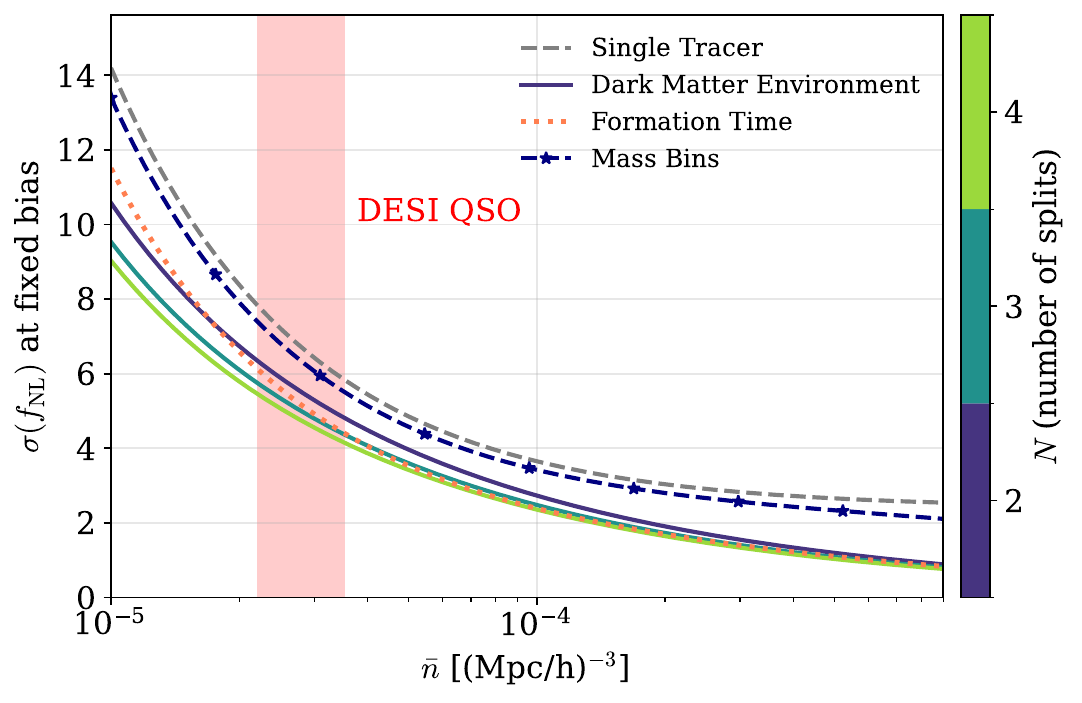}
        \caption{Forecasted constraints at the DESI quasar (QSO) effective redshift. In contrast to Fig.~\ref{fig:Fisher}, this calculation assumes a fixed bias, as the DESI quasar bias differs significantly from the halo bias at the relevant number density. Incorporating a density split yields a $\sim 20\%$ improvement relative to a single-tracer analysis.}
        \label{fig:QSOforecast}
    \end{subfigure}
\end{figure}

As discussed in Sec.~\ref{sec:TheoBack}, in the cosmic-variance-dominated limit the Fisher information on $\fnl$ from a two-tracer analysis scales as
\begin{equation}
    F_{\fnl}
    \propto \left(b_\phi^{(1)}b_\phi^{(2)}\right)^2\left( b_1^{(1)}/b_\phi^{(1)} - b_1^{(2)}/b_\phi^{(2)} \right)^2. \label{eq:AppendixF}
\end{equation}
The optimal two-tracer split should therefore maximize the difference in $b_1/b_\phi$ between the resulting subsamples, weighted by the product of the two PNG responses. To assess how closely the environmental split approaches this limit, the lower panel of Fig.~\ref{fig:FurtherForecasts} shows the first factor $b_1/b_\phi$ as a function of number density for several tracer selections. The dark-matter-environment split produces by far the largest separation in $b_1/b_\phi$, with the two subsamples evolving in opposite directions and no other pair of tracers providing a comparable lever arm. The second factor, $b_\phi^{(1)}b_\phi^{(2)}$,  is what allows a selection with a smaller lever arm to remain competitive.

It may therefore appear surprising that the environmental split does not outperform the formation-time-based analysis. There are two reasons for this. First, the PNG response increases with halo age and so the $b_\phi^{(1)}b_\phi^{(2)}$ factor in Eq.~\eqref{eq:AppendixF} is larger than for the environmentally selected tracers. Second, and more importantly, the formation-time constraining power is not primarily a multitracer effect at all. At the LRG number density, the oldest-halo subsample alone already achieves $\sigma(\fnl)\sim 4.5$, while the younger subsamples contribute comparatively little to the total constraining power.

This observation also explains why combining the dark-matter-environment split with additional selections based on halo mass or formation time yields only modest gains, as shown in the upper panel of Fig.~\ref{fig:FurtherForecasts}. The environmental split already captures nearly all of the Fisher information available through differences in $b_1/b_\phi$. At sufficiently high number density, however, the PNG response of the full sample becomes small, and selecting older halos can still provide a substantial benefit by increasing $b_\phi$. For example, at
$\bar n \sim 10^{-3}\,(h/\mathrm{Mpc})^3$,
including formation-time information improves the constraints by an additional $\sim 50\%$ relative to the environmental split alone.

Finally, we consider the applicability of our method to the DESI quasar sample. Unlike LRGs, which are thought to be reasonably well approximated as mass-selected halos and to follow the universality prediction for $b_\phi$, quasars are preferentially associated with recent mergers and are therefore expected to exhibit a weaker PNG response. We account for this effect by adopting $p=1.4$ in the relation
$b_\phi=2\delta_c(b_1-p)$~\cite{fondi2026assemblybiaslocalprimordial}.
Moreover, the observed QSO bias differs substantially from the bias of a mass-selected halo sample at the same redshift and number density. Consequently, the forecast shown in Fig.~\ref{fig:QSOforecast} is performed at fixed QSO bias as the number density is varied, rather than using the abundance-dependent halo bias adopted in Fig.~\ref{fig:Fisher}. At the DESI QSO number density, splitting the sample by its reconstructed large-scale dark-matter environment reduces the forecasted uncertainty on $\fnl$ by $\sim 20\%$. 

\newpage
\thispagestyle{empty}

\end{document}